\documentclass[letterpaper,12pt]{article}
\usepackage[toc,title,page]{appendix}
\usepackage{graphicx}
\usepackage{subfigure}
\usepackage[percent]{overpic}
\usepackage{graphics}
\usepackage{caption}
\usepackage{xcolor}
\usepackage{adjustbox}
\usepackage{tabularx} 
\usepackage{amsmath}  
\usepackage{graphicx} 
\usepackage[margin=1in,letterpaper]{geometry} 
\usepackage[final]{hyperref} 
\hypersetup{
	colorlinks=true,       
	linkcolor=blue,        
	citecolor=blue,        
	filecolor=magenta,     
	urlcolor=blue
}
\usepackage[title]{appendix}
\usepackage{filecontents}
\usepackage{natbib}
\usepackage{graphics,graphicx}
\usepackage{graphicx} 
\usepackage{epstopdf}
\usepackage{amsmath}
\usepackage{amssymb,amsmath}
\usepackage{bm}
\usepackage{amsmath}
\usepackage{booktabs}
\usepackage{appendix}
\usepackage{xcolor}

\usepackage{mathtools}
\usepackage{nicefrac}
\numberwithin{equation}{section}
\usepackage{graphicx}
\usepackage{float}
\usepackage{subfigure}
\usepackage{blindtext}
\usepackage{mathrsfs}
\usepackage{import}
\usepackage{xifthen}
\usepackage{pdfpages}
\usepackage{transparent}
\usepackage{graphicx}
\usepackage{float}
\usepackage[T1]{fontenc}
\usepackage[utf8]{inputenc}
\usepackage{authblk}
\usepackage{import}
\usepackage{xifthen}
\usepackage{pdfpages}
\usepackage{transparent}

\usepackage{setspace}
\title{\textbf{Drawing of Weakly Viscoelastic Fluid Tubes}}
\author[a]{Diandian Gu}
\author[a]{Jonathan J. Wylie}
\author[b]{Dongdong He}
\author[c]{Yvonne M. Stokes}
\affil[a]{Department of Mathematics, City University of Hong Kong, 83 Tat Chee Avenue, Kowloon Tong, Hong Kong}
\affil[b]{School of Science and Engineering,
	The Chinese University of Hong Kong, Shenzhen, Guangdong 518172, PR China}
\affil[c]{School of Mathematical Sciences and Institute for Photonics and Advanced Sensing,
	The University of Adelaide, SA 5005, Australia}

\date{}
\allowdisplaybreaks[4]
\begin{document}
\author{}
\maketitle
\begin{abstract}
	We explore the drawing of an axisymmetric viscoelastic tube subject to inertial and surface tension effects. We adopt the Giesekus constitutive model and derive
	asymptotic long-wave equations for weakly viscoelastic effects. Intuitively, one might imagine that the elastic stresses should act to prevent hole closure during the drawing process. Surprisingly, our results show that the hole closure at the outlet is enhanced by elastic effects for most parameter values. 
	However, the opposite is true if the tube has a very large hole size at the inlet of the device or if the axial stretching is very weak.
	We explain the physical mechanism underlying this phenomenon by examining how the second normal stress difference induced by elastic effects modifies the hole evolution process. We also determine how viscoelasticity affects the stability of the drawing process and show that elastic effects are always destabilizing for negligible inertia. This is in direct contrast to the case of a thread without a hole for which elastic effects are always stabilizing. On the other hand, our results show that if the inertia is non-zero, elastic effects can be either stabilizing or destabilizing depending on the parameters.
\end{abstract}

\section{Introduction}
Optical fibres form a critical component of the modern interconnected world in which information must be transmitted over extremely long distances and at high speeds in a secure and reliable way. In recent years the use of micro-structured optical fibres (MOFs) has become significant in a number of applications \citep{pal2010guided,chen2016asymptotic,liu2017microstructured}. MOFs are fibres that contain an array of air-holes that run parallel to the axis of the fibre. The air holes modify the optical transmission properties and allow for fibres with highly customizable optical properties that are extremely valuable in a number of important applications. These include micro-electrode fabrication \citep{puil2003heat,huang2007formation}, microscopy fabrication \citep{gallacchi2001well} and high-accuracy sensing \citep{xue2023review}. The optical properties are highly sensitive to small deviations in the hole geometry, and it is therefore extremely important to have a very carefully controlled fabrication process.

The most common method for fabricating fibres is by taking a relatively large preform with a hole structure and stretching it so as to reduce the cross-sectional area by factors of up to ten thousand times. For Newtonian materials, in the absence of surface tension, theoretical results have shown that the stretching process will precisely reduce the relative size of the thread whilst maintaining the geometry \citep{dewynne1994slender}. However, in real-world applications surface tension effects are non-negligible and this can cause significant deviations in the shapes and positions of the hole structures. This is particularly critical in the production of MOFs, as the optical properties are highly sensitive to the hole geometry. In some cases, some of the holes may even close. Pressurizing air channels can be used to prevent this but also causes geometry changes and may cause holes to burst \citep{chen2016asymptotic}. Therefore, one needs to take into account the effects of surface tension and pressure to ensure that the desired hole structure is achieved in the extended fibre. Another important issue in the fabrication process is that the drawing of threads is subject to an oscillatory instability that causes non-uniformities in the axial thread profiles \citep{denn1980continuous}. If severe, this will render an optical fibre not fit for purpose. Suppressing instabilities of this type is therefore a critical aspect of fibre fabrication.

While most MOFs are made from glass materials, which can be approximated as having Newtonian rheology, there has been growing interest in using polymer materials for fibre production \citep{van2001microstructured,large2008handling,argyros2013microstructures,gierej2022microstructured}. Polymer fibres offer advantages such as lower processing temperatures and ease of preform creation, and a wide range of polymers have been found suitable for the fabrication process. Unlike glass materials, polymer materials can exhibit significant viscoelasticity. This leads to the question of whether viscoelastic polymeric materials can be utilized to better achieve the manufacturing goals of preserving internal hole structures and controlling instabilities. Hence, a motivation of this work is to determine how elastic effects can be leveraged to improve the fabrication process for the drawing of `holey' fibres and other related applications.

The drawing of threads has an extensive history dating back to \cite{matovich1969spinning} who considered the drawing of a Newtonian solid thread (with no holes) and determined the criteria for instability. \cite{shah1972stability} proposed a generalized theoretical model by considering the effects of inertia, gravity, surface tension, along with thermal effects. It turns out that the interaction between the various physical effects gives rise to complicated and very rich dynamics that have been studied by a number of authors \citep{geyling1976basic,geyling1980extensional,cummings1999evolution,forest2001unsteady,wylie2007thermal,suman2009draw,taroni2013asymptotic,bechert2017combined,philippi2022linear}. All of these works considered Newtonian solid threads with no internal holes.

The dynamics of drawing solid threads composed of viscoelastic fluids have also been widely investigated. \cite{denn1975mechanics} neglected surface tension and inertia, proposed approximate equations and obtained the steady-state solutions for a thread composed of a generalized Maxwell material. \cite{fisher1976theory} extended the study by including the deformation-rate dependency of polymeric materials (White-Metzner model) and performed a linear stability analysis. \cite{jung1999stability} developed a simple approximate method for determining the stability threshold for a viscoelastic solid thread, and the mechanism for draw resonance was discussed by \cite{hyun1999draw}. \cite{park1990extensional} considered the steady drawing process of a two-phase compound fibre, in which the core is a Newtonian fluid surrounded by a sheath layer that is modelled as a weakly upper-convected Maxwell fluid. \cite{lee1995stability} studied the stability of this type of flow. \cite{gupta2017stability} considered the linear stability of a compound fibre whose core is described using the eXtended Pom-Pom (XPP) model and whose sheath layer is described using the upper-convected Maxwell (UCM) model. \cite{zhou2010thermal} investigated the steady state and linear stability of the non-isothermal drawing of viscoelastic fibres whose viscosity varies with temperature. In fact, there are many works considering various aspects of the drawing of viscoelastic threads: \cite{van2012stability} considered generalized boundary conditions, \cite{gupta2015weakly} performed a weakly nonlinear analysis, \cite{gupta2016swell} considered die swell effects, and \cite{gupta2018stability} considered non-isothermal effects. All of these studies involved fluid flows with no internal holes.\\
\indent Early works on the drawing of Newtonian threads with holes was performed by \cite{pearson1970flow1,pearson1970flow2}. Subsequently, \cite{fitt2001modeling,fitt2002mathematical} derived an asymptotic mathematical model for the drawing of axisymmetric threads with an internal hole. They showed that there are negligible leading-order pressure gradients in the radial direction and so similar mathematical techniques to those used in the case of a solid thread could be used. \cite{griffiths2007surface,griffiths2008mathematical} and \cite{stokes2014drawing} developed a general mathematical framework for analyzing non-axisymmetric threads. This was generalized to include internal pressurization of holes \citep{chen2015microstructured} and thermal effects \citep{stokes2019coupled}. To our best knowledge, there are very few studies that consider the stability of threads with internal hole structure. Only very recently, \cite{wylie2023stability} studied the stability of drawing of MOFs with internal holes. All of these works are for Newtonian fluids.\\
\indent Despite the extensive work on the drawing of Newtonian threads with holes, there has been no systematic derivation or examination of viscoelastic threads with holes. At first sight, this seems to be perplexing until one realizes that in the case of viscoelastic threads there are capillary forces acting on the inside of the holes that induce pressure gradients in the radial direction that are not present in the Newtonian case. These radial pressure gradients introduce complicated feedback mechanisms between the axial and perpendicular flows that fundamentally modify the rate at which holes close and the stability characteristics of the flow. In fact, we will show that these radial pressure gradients are crucial to understanding the role that viscoelasticity plays. From a technical viewpoint the radial pressure gradients prevent one from applying the techniques used by previous authors such as \cite{fitt2001modeling} and \cite{stokes2019coupled}. Nevertheless, we will adopt the Giesekus constitutive model \citep{giesekus1982simple} and derive asymptotic long-wave equations that allow us to determine how weakly viscoelastic effects modify the Newtonian flow. Elastic effects are known to hinder the surface-tension-driven pinching that occurs in threads without holes \citep{entov1997effect,li2003drop}. Therefore, it seems reasonable that elastic stresses should oppose any inward radial flow generated by surface tension and hence act to reduce hole closure during the drawing process. However, surprisingly, we will show that elasticity typically tends to induce more rapid hole closure. On the other hand, the opposite is true if the tube has a very thin wall at the inlet of the device or if the axial stretching is very weak. By carefully examining the expressions for the hole size at the outlet of the device we will explain how the second normal stress difference induced by elastic effects modifies the hole evolution process. Furthermore, we will determine whether elastic effects stabilize or destabilize the drawing process. We will show that if inertia is negligible then elastic effects always act to destabilize the flow. However, for non-zero inertia, elastic effects can be either stabilizing or destabilizing depending on the surface tension, inlet hole size, and parameters that describe the constitutive behaviour.\\
\indent The paper is structured as follows: Section~\ref{sec:2} presents the model formulation and derives the governing equations using the assumption of a slender tube. Section~\ref{sec:3} explores the weakly elastic limit and derives the long-wave nonlinear system that describes the drawing of an axisymmetric tube made of a viscoelastic fluid. Section~\ref{sec:3.3} discusses the steady-state profiles and explains how elasticity influences hole evolution. Section~\ref{sec:5} conducts a linear stability analysis to assess the impact of elasticity on the stability of the drawing process. Finally, section~\ref{sec:6} provides a discussion of the results.

\section{Model formulation}\label{sec:2}
\begin{figure}
	\centering
	\includegraphics[width=0.23\linewidth]{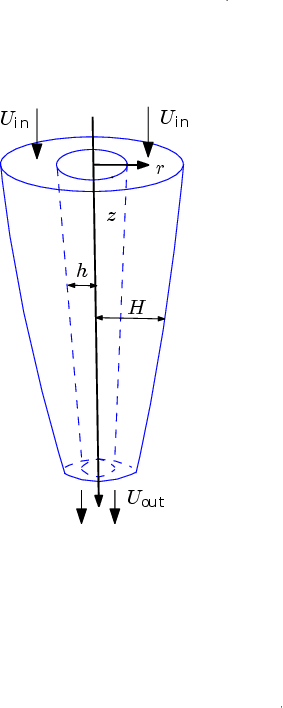}
	\caption{Schematic of the drawing process for a viscoelastic tube.}
	\label{fig:1} 
\end{figure}
We consider a slender axisymmetric tube composed of a viscoelastic incompressible fluid. This fluid is fed through an aperture of a drawing device with a constant velocity
$U_{in}$. At the aperture the outer radius is denoted by $H_{in}$, and the inner one by $h_{in}$. We define $\phi=h_{in}/H_{in}$ to be the ratio of the inner to the outer radius at the input aperture with $0<\phi<1$. At a distance $L$ from the aperture, the tube is pulled by a take-up roller such that the tube has a speed $U_{out}$. For simplicity we do not consider pressurization of the air channel, although it is straightforward to generalize. In what follows, $z$ is the distance measured along the axis of the tube, $r$ is the distance measured radially outward from the axis of the tube and $t$ is the time. The inner and outer radius of the tube are denoted by $h(z, t)$ and $H(z,t)$, respectively, as shown in Figure~\ref{fig:1}. The fluid has a velocity field $\mathbf{u}=(v, 0, u)$ in cylindrical coordinates, where $v,\ 0$ and $u$ are the velocity components in the $r,\ \theta$ and $z$ direction, respectively.

The equations for the fluid tube are given by conservation of mass and momentum:
\begin{align}
&\bm{\nabla}\cdot\mathbf{u}=0,\label{2.1}\\
&\rho(\mathbf{u}_{,t}+\mathbf{u}\cdot\bm{\nabla}\mathbf{u})=\bm{\nabla}\cdot\bm{\sigma},\label{2.2}
\end{align}
where commas are used to represent partial derivatives with respect to the letters following the commas and $\bm{\nabla}$ is the differential operator given by $\bm{\nabla}=(\partial_r, \partial_\theta/r, \partial_z)$ in the cylindrical coordinates, $\rho$ is the density, and $\bm{\sigma}$ is the total stress tensor. We assume that the viscoelasticity in the fluid arises from polymers that are dissolved in a Newtonian solvent and adopt the Giesekus model \citep{giesekus1982simple}. Therefore, the stress tensor $\bm{\sigma}$ satisfies the following constitutive relation
\begin{align}
&\bm{\sigma}=-p\mathbf{I}+2\eta_s\mathbf{D}+\bm{\tau},\label{2.3}\\
&\bm{\tau}+\lambda\overset{\triangledown}{\bm{\tau}}+\frac{\alpha \lambda}{\eta_p}(\bm{\tau})^2=2\eta_p\mathbf{D},\label{2.4}\\
&\overset{\triangledown}{\bm{\tau}}=\bm{\tau}_{,t}+\mathbf{u}\cdot \bm{\nabla}\bm{\tau}-(\bm{\nabla}\mathbf{u})^\intercal\cdot\bm{\tau}-\bm{\tau}\cdot(\bm{\nabla}\mathbf{u}),\label{2.5a}
\end{align}
where $p$ is the pressure, $\mathbf{I}$ is the identity matrix, $\mathbf{D}=(\bm{\nabla}\mathbf{u}+(\bm{\nabla}\mathbf{u})^\intercal)/2$ is the strain-rate tensor, $\eta_s$ is the viscosity of the solvent, $\eta_p$ is the viscosity of the polymer, $\bm{\tau}$ is the polymer stress tensor, and $\lambda$ is the relaxation time for the polymer. The parameter $\alpha$ is the mobility factor which is dimensionless and represents the importance of the quadratic term in the constitutive law \eqref{2.4}. In the case of $\alpha=0$,  the Giesekus model reduces to the Oldroyd-B model \citep{oldroyd1950formulation}. One well-known drawback of the Oldroyd-B model is that it can give rise to infinite stresses at finite elongational rates \citep{bird1987dynamics,tanner2000engineering,renardy2000mathematical,evans2017stresses}. The Giesekus model removes this unphysical feature by adding a term that is quadratic in the stress tensor into the constitutive relation. We note that $\alpha$ should be smaller than $0.5$ to avoid a non-monotonic dependence of the shear stress on the shear rate in simple shear flows \citep{morozov2015introduction}. We note that for $\alpha=0$ and $\eta_s=0$, the Giesekus model reduces to the upper-convected Maxwell model.

On the inner and outer surfaces of the tube, the dynamic boundary conditions
are
\begin{align}
&\mathbf{n}_1\cdot\bm{\sigma}\cdot\mathbf{n}_1=-\gamma \kappa_1,\quad \mathbf{t}_1\cdot\bm{\sigma}\cdot\mathbf{n}_1=0\quad \text{at}\quad r=h(z,t),\label{2.5}\\
&\mathbf{n}_2\cdot\bm{\sigma}\cdot\mathbf{n}_2=-\gamma \kappa_2,\quad \mathbf{t}_2\cdot\bm{\sigma}\cdot\mathbf{n}_2=0\quad \text{at}\quad r=H(z,t).\label{2.6}
\end{align}
Here, $\gamma$ is the surface tension, and $\kappa_1$, $\kappa_2$ are the mean curvatures
of inner and outer surfaces, respectively, given by
\begin{align}
\kappa_1=\bigg[\frac{h_{,zz}}{(1+h_{,z}^2)^{3/2}}-\frac{1}{h(1+h_{,z}^2)^{1/2}}\bigg],\quad \kappa_2=-\bigg[\frac{H_{,zz}}{(1+H_{,z}^2)^{3/2}}-\frac{1}{H(1+H_{,z}^2)^{1/2}}\bigg]\label{2.7}.
\end{align}
The vectors $\mathbf{n}_i$ and $\mathbf{t}_{i}$ $(i=1,\ 2)$ are the unit vectors on the inner and outer surfaces in the normal and tangential directions, which are given by
\begin{align}
&\mathbf{n}_1=\frac{1}{\sqrt{1+h_{,z}^2}}(-1,\ 0,\ h_{,z}),\quad \mathbf{n}_2=\frac{-1}{\sqrt{1+H_{,z}^2}}(-1,\ 0,\ H_{,z}),\label{2.8}\\
&\mathbf{t}_1=\frac{1}{\sqrt{1+h_{,z}^2}}(h_{,z},\ 0,\ 1),\quad 
\mathbf{t}_2=\frac{-1}{\sqrt{1+H_{,z}^2}}(H_{,z},\ 0,\ 1).\label{2.9}
\end{align}
The kinematic boundary conditions are
\begin{align}
h_{,t}+uh_{,z}=v\quad \text{at}\quad r=h(z,t),\quad H_{,t}+uH_{,z}=v\quad \text{at}\quad r=H(z,t).\label{2.10}
\end{align}
At the device inlet $z=0$, the boundary conditions are
\begin{align}
u=U_{in},\quad h=h_{in},\quad H=H_{in}.
\end{align}
At the outlet $z=L$, the boundary condition is
\begin{align}
u=U_{out}.
\end{align}

\subsection{Scaling analysis and non-dimensionalization}\label{sec:2.1}
In this work, we consider a slender axisymmetric tube in which the typical radius is far less than the length, that is, $H_{in}\ll L$ so that the dimensionless parameter $\epsilon=H_{in}\sqrt{1-\phi^2}/L$ is small. We then introduce the following scalings:
\begin{align}
&z=Lz^\prime,\quad r=\epsilon L r^\prime,\quad t=\frac{L}{U_{in}}t^\prime,\quad p=\frac{\eta_0 U_{in}}{L}p^\prime,\quad u=U_{in} u^\prime,\quad v=\epsilon U_{in} v^\prime,\nonumber\\
&\tau_{rr}=\frac{\eta_0 U_{in}}{L}\tau^\prime_{rr},\quad \tau_{\theta\theta}=\frac{\eta_0 U_{in}}{L}\tau^\prime_{\theta\theta},\quad \tau_{zz}=\frac{\eta_0 U_{in}}{L}\tau^\prime_{zz},\quad \tau_{rz}=\frac{\eta_0 U_{in}}{\epsilon L}\tau^\prime_{rz},\label{2.11}
\end{align}
where $\eta_0=\eta_s+\eta_p$ is the bulk viscosity and $\tau_{rr},\ \tau_{\theta\theta},\ \tau_{zz},$ and $\tau_{rz}$ are the components of the polymer stress tensor $\bm{\tau}$. Substituting the scalings \eqref{2.11} into \eqref{2.1}$-$\eqref{2.5a} and omitting the primes for convenience, we obtain
\begin{align}
&u_{,z}+\frac{1}{r}(rv)_{,r}=0,\label{2.12}\\
&\mathrm{Re}(u_{,t}+vu_{,r}+uu_{,z})=-p_{,z}+\frac{\beta}{\epsilon^2}\frac{1}{r}(ru_{,r})_{,r}+\frac{\beta}{r}(rv_{,z})_{,r}+2\beta u_{,zz}+\frac{1}{\epsilon^2}\frac{1}{r}(r\tau_{rz})_{,r}+\tau_{zz,z},\label{2.13}\\
&\mathrm{Re}(v_{,t}+v v_{,r}+u v_{,z})=-\frac{1}{\epsilon^2}p_{,r}+\frac{\beta}{\epsilon^2}\big(\frac{1}{r}(rv)_{,r}\big)_{,r}+\beta v_{,zz}+\frac{1}{\epsilon^2}\big(\tau_{rr,r}+\tau_{rz,z}+\frac{\tau_{rr}-\tau_{\theta\theta}}{r}\big),\label{2.14}\\
&\tau_{rr}+\mathrm{De}\big(\tau_{rr,t}+v\tau_{rr,r}+u\tau_{rr,z}-2\tau_{rr}v_{,r}-2\tau_{rz}v_{,z}\big)+\frac{\alpha \mathrm{De}}{1-\beta}\big(\tau_{rr}^2+\frac{1}{\epsilon^2}\tau_{rz}^2\big)=2(1-\beta)v_{,r},\label{2.15}\\
&\tau_{\theta\theta}+\mathrm{De}\big(\tau_{\theta\theta,t}+v\tau_{\theta\theta,r}+u\tau_{\theta\theta,z}-2\tau_{\theta\theta}\frac{v}{r}\big)+\frac{\alpha \mathrm{De}}{1-\beta}\tau_{\theta\theta}^2=2(1-\beta)\frac{v}{r},\label{2.16}\\
&\tau_{zz}+\mathrm{De}\big(\tau_{zz,t}+v\tau_{zz,r}+u\tau_{zz,z}-\frac{2}{\epsilon^2}\tau_{rz}u_{,r}-2\tau_{zz}u_{,z}\big)+\frac{\alpha \mathrm{De}}{1-\beta}\big(\tau_{zz}^2+\frac{1}{\epsilon^2}\tau_{rz}^2\big)=2(1-\beta)u_{,z},\label{2.17}\\
&\tau_{rz}+\mathrm{De}\big(\tau_{rz,t}+v\tau_{rz,r}+u\tau_{rz,z}+\tau_{rz}\frac{v}{r}-\tau_{rr}u_{,r}-\epsilon^2\tau_{zz}v_{,z}\big)+\frac{\alpha \mathrm{De}}{1-\beta}(\tau_{rr}+\tau_{zz})\tau_{rz}\nonumber\\
&\qquad\qquad\qquad\qquad\qquad\qquad\qquad\qquad\qquad\qquad\qquad\qquad=(1-\beta)(\epsilon^2 v_{,z}+u_{,r}),\label{2.18}
\end{align}
where 
\begin{align}
\beta=\frac{\eta_s}{\eta_0},\quad \mathrm{De}=\frac{\lambda U_{in}}{L},\quad \mathrm{Re}=\frac{\rho U_{in} L}{\eta_0}.\label{2.19}
\end{align}
The parameter $\beta$ is the solvent to bulk viscosity ratio. The Deborah number $\mathrm{De}$ is the ratio of the relaxation time to the characteristic time for fluid to flow through the device. The Reynolds number $\mathrm{Re}$ quantifies the relative importance of inertial and viscous effects. We note that if $\mathrm{De}=0$, equations \eqref{2.12}$-$\eqref{2.14} reduce to those for the Newtonian flow, with viscosity $\eta_0=\eta_s+\eta_p$. 

By substituting \eqref{2.11} into the normal components of the dynamic boundary conditions in \eqref{2.5}$-$\eqref{2.6}, we obtain
\begin{align}
&p-\frac{2\beta}{1+\epsilon^2h_{,z}^2}\big[v_{,r}-h_{,z}(\epsilon^2 v_{,z}+u_{,r})+\epsilon^2 h_{,z}^2u_{,z}\big]-\frac{1}{1+\epsilon^2h_{,z}^2}\big(\epsilon^2h_{,z}^2 \tau_{zz}-2h_{,z}\tau_{rz}+\tau_{rr}\big)\nonumber\\&=\frac{1}{\mathrm{Ca}}\bigg[\frac{\epsilon^2 h_{,zz}}{(1+\epsilon^2h_{,z}^2)^{3/2}}-\frac{1}{h(1+\epsilon^2h_{,z}^2)^{1/2}}\bigg],\label{2.20}\\
&p-\frac{2\beta}{1+\epsilon^2H_{,z}^2}\big[v_{,r}-H_{,z}(\epsilon^2 v_{,z}+u_{,r})+\epsilon^2 H_{,z}^2u_{,z}\big]-\frac{1}{1+\epsilon^2H_{,z}^2}\big(\epsilon^2H_{,z}^2 \tau_{zz}-2H_{,z}\tau_{rz}+\tau_{rr}\big)\nonumber\\
&=-\frac{1}{\mathrm{Ca}}\bigg[\frac{\epsilon^2 H_{,zz}}{(1+\epsilon^2H_{,z}^2)^{3/2}}-\frac{1}{H(1+\epsilon^2H_{,z}^2)^{1/2}}\bigg],\label{2.21}
\end{align}
where 
\begin{align}
\mathrm{Ca}=\frac{\eta_0 U_{in} \epsilon}{\gamma},
\end{align}
which denotes the capillary number and quantifies the relative importance of viscous and surface tension forces. The non-dimensional tangential components of the dynamic boundary conditions are
\begin{align}
&2\beta\epsilon^2h_{,z}(v_{,r}-u_{,z})+\beta\big(1-\epsilon^2h_{,z}^2\big)(\epsilon^2v_{,z}+u_{,r})+\epsilon^2h_{,z}(\tau_{rr}-\tau_{zz})+\big(1-\epsilon^2h_{,z}^2\big)\tau_{rz}=0,\label{2.22}\\
&2\beta\epsilon^2H_{,z}(v_{,r}-u_{,z})+\beta\big(1-\epsilon^2H_{,z}^2\big)(\epsilon^2v_{,z}+u_{,r})+\epsilon^2H_{,z}(\tau_{rr}-\tau_{zz})+\big(1-\epsilon^2H_{,z}^2\big)\tau_{rz}=0.\label{2.23}
\end{align}
The kinematic boundary conditions \eqref{2.10} remain unchanged under the scaling transformations. The dimensionless boundary conditions at the device inlet and outlet are given by
\begin{align}
&u=1,\quad h=\frac{\phi}{\sqrt{1-\phi^2}},\quad H=\frac{1}{\sqrt{1-\phi^2}}\quad \text{at}\quad z=0,\label{2.27a}\\
&u=\mathrm{D}\quad \text{at}\quad z=1,\label{2.28a}
\end{align}
where $\mathrm{D}$ is the draw ratio, given by
\begin{align}
\mathrm{D}=\frac{U_{out}}{U_{in}}.
\end{align}

Fibre drawing is a process that is performed using devices of very different sizes and using materials with dramatically different viscosities. This means that one typically has to consider a wide range of values for many of the parameters. The values of parameters within such flows can vary significantly based on the specific nature of the industrial process. Typically, the value of $\epsilon$ remains relatively small, usually not exceeding $\mathcal{O}(10^{-1})$ but often much smaller. In scenarios involving drawing, the parameter $\mathrm{D}$ can reach values on the scale of $\mathcal{O}(10^3)$ or higher. Whereas extrusion flows have values of $\mathrm{D}$ that are typically $\mathcal{O}(1)$ \citep{tronnolone2017extrusion}. Furthermore, both small and large Reynolds and capillary numbers are important \citep{bechert2017combined}. Considering these factors, we will explore a wide spectrum of potential parameter values in this paper.

Due to the fact that $\epsilon$ only appears in the governing equations and boundary conditions as $\epsilon^2$, we proceed by proposing the following asymptotic expansions and assuming that $\mathrm{Re},\ \mathrm{Ca},\ \alpha$, and $\beta$ are $\mathcal{O}(1)$ quantities,
\begin{align}
&\mathbf{u}=\mathbf{u_0}+\epsilon^2 \mathbf{u_2}+\mathcal{O}(\epsilon^4),\label{2.24}\\
&\bm{\tau}=\bm{\tau_{0}}+\epsilon^2\bm{\tau_{2}}+\mathcal{O}(\epsilon^4),\label{2.25}\\
&p=p_{0}+\epsilon^2 p_2+\mathcal{O}(\epsilon^4).\label{2.26}
\end{align}
Substituting \eqref{2.24}$-$\eqref{2.26} into \eqref{2.12}$-$\eqref{2.23} and equating the powers of $\epsilon$, we obtain a set of partial differential equations and boundary conditions for each power of $\epsilon$. We are interested in the leading-order asymptotic behaviour. From \eqref{2.13} and \eqref{2.22}$-$\eqref{2.23}, we obtain
\begin{align}
u_{0,r}=0,\quad \tau_{rz0}=0.\label{2.27}
\end{align}
Hence, $u_0$ is independent of $r$, and after integrating the leading-order version of the continuity equation \eqref{2.12}, we obtain
\begin{align}
v_0=-\frac{r}{2}u_{0,z}+\frac{C(z,t)}{r},\label{2.28}
\end{align}
where $C(z,t)$ is an integration function that will be determined later using the boundary conditions.\\
The leading-order momentum equations are given by
\begin{align}
&\mathrm{Re}(u_{0,t}+u_0u_{0,z})=-p_{0,z}+\frac{\beta}{r}(ru_{2,r})_{,r}+\frac{\beta}{r}(rv_{0,z})_{,r}+2\beta u_{0,zz}+\frac{1}{r}(r\tau_{rz2})_{,r}+\tau_{zz0,z},\label{2.29}\\
&\tau_{rr0,r}+\frac{\tau_{rr0}-\tau_{\theta\theta0}}{r}-p_{0,r}=0.\label{2.30}
\end{align}
The leading-order constitutive equations are 
\begin{align}
&\tau_{rr0}+\mathrm{De}\big(\mathscr{L}(\tau_{rr0})-2\tau_{rr0}v_{0,r}\big)+\frac{\alpha \mathrm{De}}{1-\beta}\tau_{rr0}^2=2(1-\beta)v_{0,r},\label{2.31}\\
&\tau_{\theta\theta 0}+\mathrm{De}\big(\mathscr{L}(\tau_{\theta\theta 0})-2\tau_{\theta\theta 0}\frac{v_0}{r}\big)+\frac{\alpha \mathrm{De}}{1-\beta}\tau_{\theta\theta 0}^2=2(1-\beta)\frac{v_0}{r},\label{2.32}\\
&\tau_{zz0}+\mathrm{De}\big(\mathscr{L}(\tau_{zz0})-2\tau_{zz0}u_{0,z}\big)+\frac{\alpha \mathrm{De}}{1-\beta}\tau_{zz0}^2=2(1-\beta)u_{0,z},\label{2.33}
\end{align}
where $\mathscr{L}(\cdot)=(\cdot)_{,t}+v_0(\cdot)_{,r}+u_0(\cdot)_{,z}$. The leading-order normal components of the dynamic boundary conditions are
\begin{align}
p_0-2\beta v_{0,r}-\tau_{rr0}=-\frac{1}{\mathrm{Ca} h}\quad \text{at}\quad r=h(z,t),\label{2.34}\\
p_0-2\beta v_{0,r}-\tau_{rr0}=\frac{1}{\mathrm{Ca} H}\quad \text{at}\quad r=H(z,t).\label{2.35}
\end{align}
Considering $\mathcal{O}(\epsilon^2)$ terms of the tangential components of the dynamic boundary conditions \eqref{2.22}$-$\eqref{2.23} gives
\begin{align}
&2\beta h_{,z}(v_{0,r}-u_{0,z})+\beta u_{2,r}+\beta v_{0,z} +h_{,z}(\tau_{rr0}-\tau_{zz0})+\tau_{rz2}=0\quad \text{at}\quad r=h(z,t),\label{2.36}\\
&2\beta H_{,z}(v_{0,r}-u_{0,z})+\beta u_{2,r}+\beta v_{0,z} +H_{,z}(\tau_{rr0}-\tau_{zz0})+\tau_{rz2}=0\quad \text{at}\quad r=H(z,t).\label{2.37}
\end{align}
To obtain a long-wave equation, we multiply \eqref{2.29} by $r$, integrate with respect to $r$, and use \eqref{2.36}$-$\eqref{2.37} to obtain
\begin{align}
\frac{1}{2}\mathrm{Re}(u_{0,t}+u_0 u_{0,z})(H^2-h^2)=&-\int_{h}^{H}r p_{0,z}dr+\beta(H^2-h^2)u_{0,zz}+\int_{h}^{H}r\tau_{zz0,z}dr\nonumber\\
&+2\beta h h_{,z}(v_{0,r}|_{h}-u_{0,z})-2\beta H H_{,z}(v_{0,r}|_{H}-u_{0,z})\nonumber\\
&+hh_{,z}(\tau_{rr0}-\tau_{zz0})|_{h}-HH_{,z}(\tau_{rr0}-\tau_{zz0})|_{H},\label{2.38}
\end{align}
where `$|_{h}$' and `$|_{H}$' denote evaluation at the $r=h$ and $r=H$ boundaries, respectively. Moreover, by substituting \eqref{2.28} into the kinematic boundary conditions \eqref{2.10}, we obtain the equations for $h$ and $H$,
\begin{align}
&(h^2)_{,t}+(h^2u_0)_{,z}=2C(z,t),\label{2.39}\\
&(H^2)_{,t}+(H^2 u_0)_{,z}=2C(z,t).\label{2.40}
\end{align}
Consequently, by using \eqref{2.28} to eliminate $v_0$ we obtain a closed system, \eqref{2.30}$-$\eqref{2.35} and \eqref{2.38}$-$\eqref{2.40}, for $u_0,\ h,\ H,\ p_0,\ C,\ \tau_{rr0},\ \tau_{zz0},$ and $\tau_{\theta\theta0}$. 

In the case of a Newtonian fluid, it can be shown that the leading-order pressure $p_0$ is independent of $r$ \citep{fitt2002mathematical}. This allows for a straightforward evaluation of the integrals in \eqref{2.38}, resulting in a set of three equations for $u_0$, $h$, and $H$ that are independent of $r$. However, in the current problem the pressure is not independent of $r$ and it turns out that the integrals in \eqref{2.38} cannot be expressed in terms of elementary functions, so one cannot directly derive long-wave equations that are tractable. In the case of solid viscoelastic threads without internal holes, many authors have proceeded by expanding the various dynamic variables as Taylor series in powers of $\epsilon r$ \citep{eggers2008physics,li2021dynamics}. This approach is suitable because the point $r=0$ is included in the domain, so the regularity conditions at $r=0$ preclude terms that are singular. However, in the current problem, $r=0$ is not part of the domain, which means that terms arising from the $C/r$ term in \eqref{2.28} cannot be accurately described using a Taylor series expansion. In fact, if one attempts to expand in powers of $\epsilon r$, one immediately sees that such an expansion cannot be compatible with the normal boundary conditions \eqref{2.34}$-$\eqref{2.35}. Therefore, an alternative method is needed to derive the long-wave equations.

\section{Low Deborah number asymptotics }\label{sec:3}
In this section, we will simplify our system of equations and boundary conditions, \eqref{2.30}$-$\eqref{2.35} and \eqref{2.38}$-$\eqref{2.40}, using the small Deborah number assumption. On examination of \eqref{2.28}, \eqref{2.30}, \eqref{2.31}, and \eqref{2.33}, we observe that the elastic stresses depend on $r$ and are strongly coupled with the pressure $p_0$ and velocity $u_0$. This makes it challenging to deal with the integrals in \eqref{2.38}. In order to overcome this challenge and obtain long-wave equations, we consider the weakly viscoelastic limit, $\mathrm{De}\ll 1$. Actually, many works consider the weakly viscoelastic limit for various problems and a broad range of important results have been found using such techniques \citep{zhou2016viscoelastic,boyko2022pressure,boyko2023non,datt2019note,de2015locomotion,binagia2020swimming}. Following their methodology, we pose an expansion of the form
\begin{align}
\psi=\psi^{(0)}+\mathrm{De}\psi^{(1)}+\mathcal{O}(\mathrm{De}^2),\label{3.1}
\end{align}
where $\psi$ represents the quantities $\bm{\tau}_0,\ p_0,\ \bm{u}_0,\ h,\ H$ and $C$.

\subsection{Leading-order long-wave equations in $\mathrm{De}$}
Substituting expressions of the form \eqref{3.1} into \eqref{2.28}, at leading order in $\mathrm{De}$ we obtain
\begin{align}
v^{(0)}_0=-\frac{r}{2}u^{(0)}_{0,z}+\frac{C^{(0)}(z,t)}{r}.\label{3.2}
\end{align}
Similar substitutions into \eqref{2.31}$-$\eqref{2.33} give
\begin{align}
&\tau^{(0)}_{rr0}=2(1-\beta)v^{(0)}_{0,r},\label{3.4}\\
&\tau^{(0)}_{\theta\theta 0}=2(1-\beta)\frac{v^{(0)}_0}{r},\label{3.5}\\
&\tau^{(0)}_{zz0}=2(1-\beta)u^{(0)}_{0,z},\label{3.6}
\end{align}
while \eqref{2.30} gives, using \eqref{3.2}$-$\eqref{3.5},
\begin{align}
p^{(0)}_{0,r}=0.\label{3.7}
\end{align}
Thus $p^{(0)}_0=A^{(0)}(z,t)$ is independent of $r$, where the integration function $A^{(0)}$ must be determined using the boundary conditions.

Substituting \eqref{3.1} into \eqref{2.34}$-$\eqref{2.35}, we find that the equations at leading order in $\mathrm{De}$ are an algebraic system in terms of $A^{(0)}$ and $C^{(0)}$, whose solutions are
\begin{align}
&A^{(0)}=\frac{1}{\mathrm{Ca}(H^{(0)}-h^{(0)})}-u^{(0)}_{0,z},\label{3.8}\\
&C^{(0)}=-\frac{h^{(0)} H^{(0)}}{2 \mathrm{Ca}(H^{(0)}-h^{(0)})}.\label{3.9}
\end{align}
Substituting \eqref{3.1} into \eqref{2.38}, considering the leading-order terms in $\mathrm{De}$, and using \eqref{3.2} and \eqref{3.4}$-$\eqref{3.9}, we obtain
\begin{align}
\mathrm{Re}(H^{(0)2}-h^{(0)2}) (u^{(0)}_{0,t}+u^{(0)}_0u^{(0)}_{0,z})=\frac{1}{\mathrm{Ca}}(h^{(0)}_{,z}+H^{(0)}_{,z})+3\big[(H^{(0)2}-h^{(0)2})u^{(0)}_{0,z}\big]_{,z}.\label{3.10}
\end{align}
In addition, at leading order in $\mathrm{De}$, \eqref{2.39} and \eqref{2.40} give
\begin{align}
&(h^{(0)2})_{,t}+(h^{(0)2}u^{(0)}_0)_{,z}=2C^{(0)},\label{3.11}\\
&(H^{(0)2})_{,t}+(H^{(0)2}u^{(0)}_0)_{,z}=2C^{(0)}.\label{3.12}
\end{align}
Substituting \eqref{3.1} into \eqref{2.27a}$-$\eqref{2.28a}, we obtain boundary conditions for $u^{(0)}_0,\ h^{(0)}$ and $H^{(0)}$, namely,
\begin{align}
u^{(0)}_0&=1,\quad h^{(0)}=\frac{\phi}{\sqrt{1-\phi^2}},\quad H^{(0)}=\frac{1}{\sqrt{1-\phi^2}}\quad \text{at}\quad z=0,\label{3.13}\\
u^{(0)}_0&=\mathrm{D}\quad \text{at}\quad z=1.\label{3.14}
\end{align}
The leading-order long-wave equations \eqref{3.10}$-$\eqref{3.14} are equivalent to setting $\mathrm{De}=0$ and are therefore the same as the equations for a Newtonian fluid tube with viscosity $\eta_0$. The system is therefore essentially equivalent to that derived in \cite{fitt2002mathematical}. To include elastic effects in our model we must consider the next-order terms in $\mathrm{De}$.

\subsection{First-order long-wave equations in $\mathrm{De}$}
Taking the first-order terms in $\mathrm{De}$ of \eqref{2.28}, we obtain 
\begin{align}
v^{(1)}_0=-\frac{r}{2}u^{(1)}_{0,z}+\frac{C^{(1)}(z,t)}{r}.\label{3.3}
\end{align}
The constitutive relations \eqref{2.31}$-$\eqref{2.33} yield
\begin{align}
&\tau^{(1)}_{rr0}=2(1-\beta)v^{(1)}_{0,r}-\big(\mathscr{L}^{(0)}(\tau^{(0)}_{rr0})-2\tau^{(0)}_{rr0}v^{(0)}_{0,r}\big)-\frac{\alpha}{1-\beta}\tau^{(0)2}_{rr0},\label{3.15}\\
&\tau^{(1)}_{\theta\theta 0}=2(1-\beta)\frac{v^{(1)}_0}{r}-\big(\mathscr{L}^{(0)}(\tau^{(0)}_{\theta\theta 0})-2\tau^{(0)}_{\theta\theta 0}\frac{v^{(0)}_{0}}{r}\big)-\frac{\alpha}{1-\beta}\tau^{(0)2}_{\theta\theta 0},\label{3.16}\\
&\tau^{(1)}_{zz0}=2(1-\beta)u^{(1)}_{0,z}-\big(\mathscr{L}^{(0)}(\tau^{(0)}_{zz0})-2\tau^{(0)}_{zz0}u^{(0)}_{0,z}\big)-\frac{\alpha}{1-\beta}\tau^{(0)2}_{zz0},\label{3.17}
\end{align}
where  $\mathscr{L}^{(0)}(\cdot)=(\cdot)_{,t}+v^{(0)}_0(\cdot)_{,r}+u^{(0)}_0(\cdot)_{,z}$. Using \eqref{2.30}, we obtain
\begin{align}
p^{(1)}_{0,r}=\frac{1}{r}\big(r \tau^{(1)}_{rr0}\big)_{,r}-\frac{1}{r}\tau^{(1)}_{\theta\theta 0}.\label{3.18}
\end{align}
Integrating \eqref{3.18} with respect to $r$ yields
\begin{align}
p^{(1)}_0=\int \frac{1}{r}\big(r \tau^{(1)}_{rr0}\big)_{,r} dr- \int \frac{1}{r}\tau^{(1)}_{\theta\theta 0} dr +A^{(1)}(z,t),\label{3.19}
\end{align}
where $A^{(1)}(z,t)$ is an integration function that will be determined later using the first-order normal boundary conditions that are given by
\begin{align}
&p^{(1)}_0-2\beta v^{(0)}_{0,rr}h^{(1)}-2\beta v^{(1)}_{0,r}-\tau^{(0)}_{rr0,r}h^{(1)}-\tau^{(1)}_{rr0}=\frac{h^{(1)}}{\mathrm{Ca} h^{(0)2}}\quad \text{at}\quad r=h^{(0)},\label{3.20}\\
&p^{(1)}_0-2\beta v^{(0)}_{0,rr}H^{(1)}-2\beta v^{(1)}_{0,r}-\tau^{(0)}_{rr0,r}H^{(1)}-\tau^{(1)}_{rr0}=-\frac{H^{(1)}}{\mathrm{Ca} H^{(0)2}}\quad \text{at}\quad r=H^{(0)}.\label{3.21}
\end{align}
We note that \eqref{3.20} and \eqref{3.21} form an algebraic system for $A^{(1)}(z,t)$ and $C^{(1)}(z,t)$, which can be solved to give
\begin{align}
A^{(1)}(z,t)=&-u^{(1)}_{0,z}+\frac{h^{(1)}-H^{(1)}}{\mathrm{Ca}(H^{(0)}-h^{(0)})^2}+\frac{2(1-\beta)C^{(0)2}}{h^{(0)2}H^{(0)2}}+(1-\beta)\Big[(1-\alpha)u^{(0)2}_{0,z}+u^{(0)}_{0,zt}\nonumber\\
&+u^{(0)}_0u^{(0)}_{0,zz}\Big],\label{3.25}\\
C^{(1)}(z,t)=&\frac{h^{(0)2}H^{(1)}-H^{(0)2}h^{(1)}}{2\mathrm{Ca}(H^{(0)}-h^{(0)})^2}+(\beta-1)\bigg[\frac{(h^{(0)2}+H^{(0)2})C^{(0)2}}{h^{(0)2}H^{(0)2}}-(C^{(0)}_{,t}+u^{(0)}_0C^{(0)}_{,z})\nonumber\\
&+(2\alpha-3)C^{(0)}u^{(0)}_{0,z}\bigg].\label{3.26}
\end{align}
Hence, we can use \eqref{3.3}$-$\eqref{3.17} and \eqref{3.25}$-$\eqref{3.26} to express $v^{(1)}_0,\ \tau^{(1)}_{rr0},\ \tau^{(1)}_{\theta\theta0},\ \tau^{(1)}_{zz0},\ A^{(1)}$ and $C^{(1)}$ in terms of the leading-order quantities $u^{(0)}_0,\ h^{(0)},\ H^{(0)}$ and $C^{(0)}$. Although the physical meaning of most variables is clear, the physical interpretation of $C(z,t)=C^{(0)}(z,t)+\mathrm{De}\ C^{(1)}(z,t)$ warrants some discussion. It is readily shown from \eqref{3.9} and \eqref{3.26} that $C$, which first appeared in \eqref{2.28}, will be identically zero in the case of zero surface tension ($\mathrm{Ca}=\infty$). In fact, the radial velocity $v_0$ in \eqref{2.28} is composed of two parts: $-r u_{0,z}/2$, which arises from axial stretching and mass conservation, and $C/r$, which arises from surface tension acting on the inner and outer boundaries. Thus, the quantity $C$ represents the strength of the radial flow induced by surface tension.

Considering the terms of $\mathcal{O}(\mathrm{De})$ in the axial momentum equation \eqref{2.38}, we get integrals that are straightforward to evaluate and obtain
\begin{align}
&\frac{1}{2}\mathrm{Re}(u^{(1)}_{0,t}+u^{(1)}_0 u^{(0)}_{0,z}+u^{(0)}_0 u^{(1)}_{0,z})(H^{(0)2}-h^{(0)2})=-\mathrm{Re} (u^{(0)}_{0,t}+ u^{(0)}_0u^{(0)}_{0,z})(H^{(0)}H^{(1)}-h^{(0)}h^{(1)})\nonumber\\
&-A^{(0)}_{,z}(H^{(0)}H^{(1)}-h^{(0)}h^{(1)})+2(2\alpha-1)(\beta-1)C^{(0)}C^{(0)}_{,z}\bigg(\frac{1}{H^{(0)2}}-\frac{1}{h^{(0)2}}\bigg)\nonumber\\
&+\big(\beta u^{(1)}_{0,zz}+\frac{1}{2}\tau^{(1)}_{zz0,z}-\frac{1}{2}A^{(1)}_{,z}\big)(H^{(0)2}-h^{(0)2})+\big(2\beta u^{(0)}_{0,zz}+\tau^{(0)}_{zz0,z}\big)(H^{(0)}H^{(1)}-h^{(0)}h^{(1)})\nonumber\\
&+\Big[2\beta(v^{(0)}_{0,r}-u^{(0)}_{0,z})+(\tau^{(0)}_{rr0}-\tau^{(0)}_{zz0})\Big]\Big|_{h^{(0)}}(h^{(1)}h^{(0)}_{,z}+h^{(0)}h^{(1)}_{,z})\nonumber\\
&+\Big[2\beta(v^{(1)}_{0,r}-u^{(1)}_{0,z})+(\tau^{(1)}_{rr0}-\tau^{(1)}_{zz0})\Big]\Big|_{h^{(0)}}h^{(0)}h^{(0)}_{,z}\nonumber\\
&-\Big[2\beta(v^{(1)}_{0,r}-u^{(1)}_{0,z})+(\tau^{(1)}_{rr0}-\tau^{(1)}_{zz0})\Big]\Big|_{H^{(0)}}H^{(0)}H^{(0)}_{,z}\nonumber\\
&-\Big[2\beta(v^{(0)}_{0,r}-u^{(0)}_{0,z})+(\tau^{(0)}_{rr0}-\tau^{(0)}_{zz0})\Big]\Big|_{H^{(0)}}(H^{(1)}H^{(0)}_{,z}+H^{(0)}H^{(1)}_{,z}).\label{3.27}
\end{align}
Considering the terms of $\mathcal{O}(\mathrm{De})$ in \eqref{2.39}$-$\eqref{2.40}, we obtain
\begin{align}
&2(h^{(0)} h^{(1)})_{,t}+2(h^{(0)} h^{(1)} u^{(0)}_0)_{,z}+(h^{(0)2} u^{(1)}_0)_{,z}=2C^{(1)},\label{3.28}\\
&2(H^{(0)} H^{(1)})_{,t}+2(H^{(0)} H^{(1)} u^{(0)}_0)_{,z}+(H^{(0)2} u^{(1)}_0)_{,z}=2C^{(1)}.\label{3.29}
\end{align}
Substituting \eqref{3.1} into \eqref{2.27a}$-$\eqref{2.28a}, we obtain boundary conditions for $u^{(1)}_0$, $h^{(1)}$, and $H^{(1)}$, namely
\begin{align}
u^{(1)}_0&=0,\quad h^{(1)}=0,\quad H^{(1)}=0\quad \text{at}\quad z=0,\label{3.30}\\
u^{(1)}_0&=0\quad \text{at}\quad z=1.\label{3.31}
\end{align}
As mentioned above, \eqref{3.9}$-$\eqref{3.14} represent the long-wave equations for a Newtonian fluid. We must first solve these equations to obtain $u^{(0)}_0,\ h^{(0)}$, and $H^{(0)}$, after which we substitute these solutions into \eqref{3.3}, \eqref{3.15}, \eqref{3.17}, and \eqref{3.25}$-$\eqref{3.31}, which represent equations for $u^{(1)}_0,\ h^{(1)}$, and $H^{(1)}$. The quantities $u^{(1)}_0,\ h^{(1)}$, and $H^{(1)}$ capture the leading-order effects of elastic stresses and describe the variation from the zero Deborah number flow. We note that in the limit $\beta\to1$, the quantities $u^{(1)}_0,\ h^{(1)}$, and $H^{(1)}$ tend to zero. This is because the limit $\beta\to 1$ corresponds to the case $\eta_p\to0$ that represents no polymeric effect on the flow, which is therefore purely Newtonian.

\section{Numerical method for steady-state equations}\label{sec:3.3}
In this section, we consider the steady state by setting $\partial_{t}\equiv0$ in \eqref{3.9}$-$\eqref{3.14} and \eqref{3.27}$-$\eqref{3.31}. In order to numerically obtain the steady-state solutions, we note that \eqref{3.10} is a second-order ordinary differential equation (ODE) for the quantity $u^{(0)}_0$, and \eqref{3.11}$-$\eqref{3.12} are two first-order
ODEs for $h^{(0)}$ and $H^{(0)}$. We also have three boundary conditions at $z = 0$ and one boundary condition at $z = 1$. Therefore, this system can be readily solved using a shooting method in which one starts with a guessed value of $u^{(0)}_{0,z}$ at $z = 0$, then solves \eqref{3.10}$-$\eqref{3.12} numerically
using a standard `initial' value ODE solver (e.g. \textmd{MATLAB} function `ode45') subject to the `initial' conditions \eqref{3.13}, and employs a root-finding technique (e.g. \textmd{MATLAB} function `fzero') to find $u^{(0)}_{0,z}$ at $z = 0$ such that the condition \eqref{3.14} is satisfied. Similarly, \eqref{3.27}$-$\eqref{3.29} are solved by guessing the value of $u^{(1)}_{0,z}$ and employing a root-finding technique to achieve \eqref{3.31}. Consequently, we solve $u_0,\ h$, and $H$ numerically up to $\mathcal{O}(\mathrm{De})$.

\subsection{Results of steady state }\label{sec:4.1}
\begin{figure}
	\centering
	\subfigure {\
		\begin{minipage}[b]{.46\linewidth}
			\centering
			\begin{overpic}[scale=0.44]{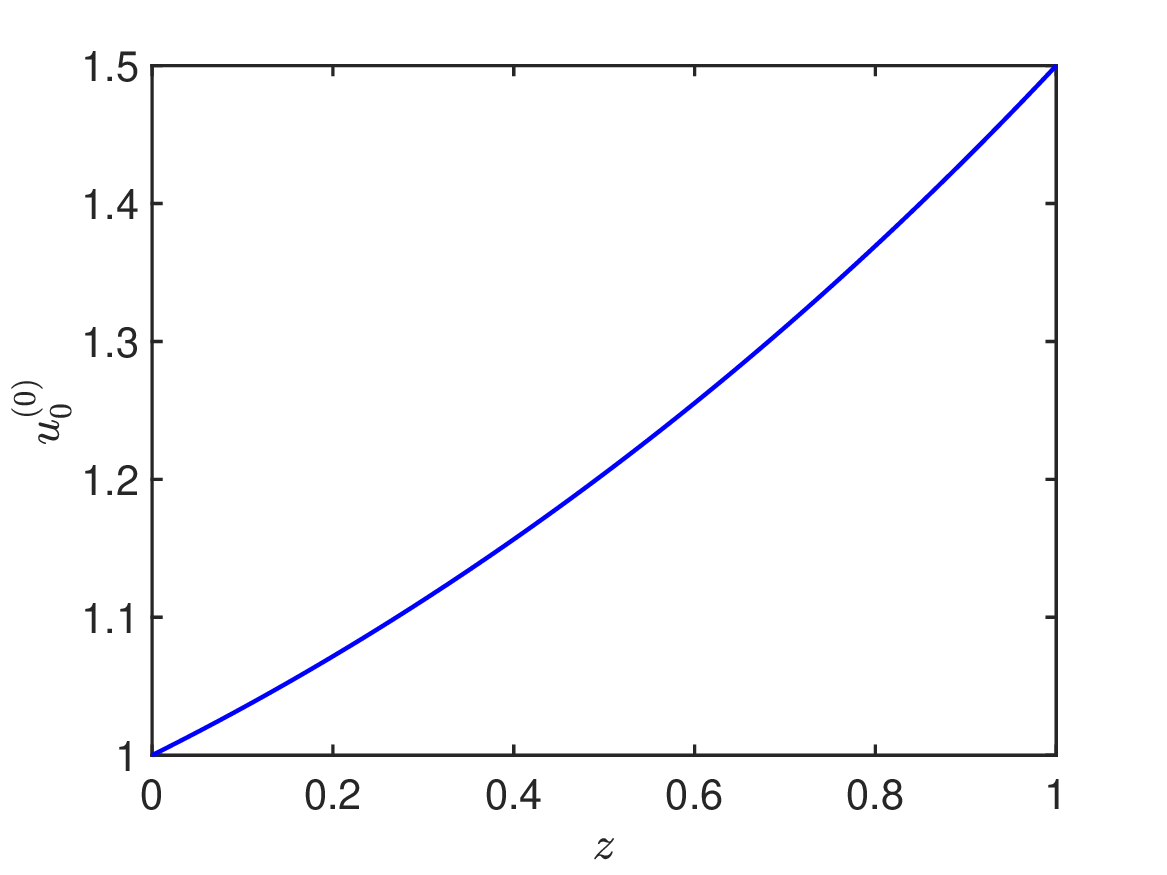}
				\put(3,72){(a)}
			\end{overpic}
		\end{minipage}
	}
	\subfigure {\
		\begin{minipage}[b]{.46\linewidth}
			\centering
			\begin{overpic}[scale=0.44]{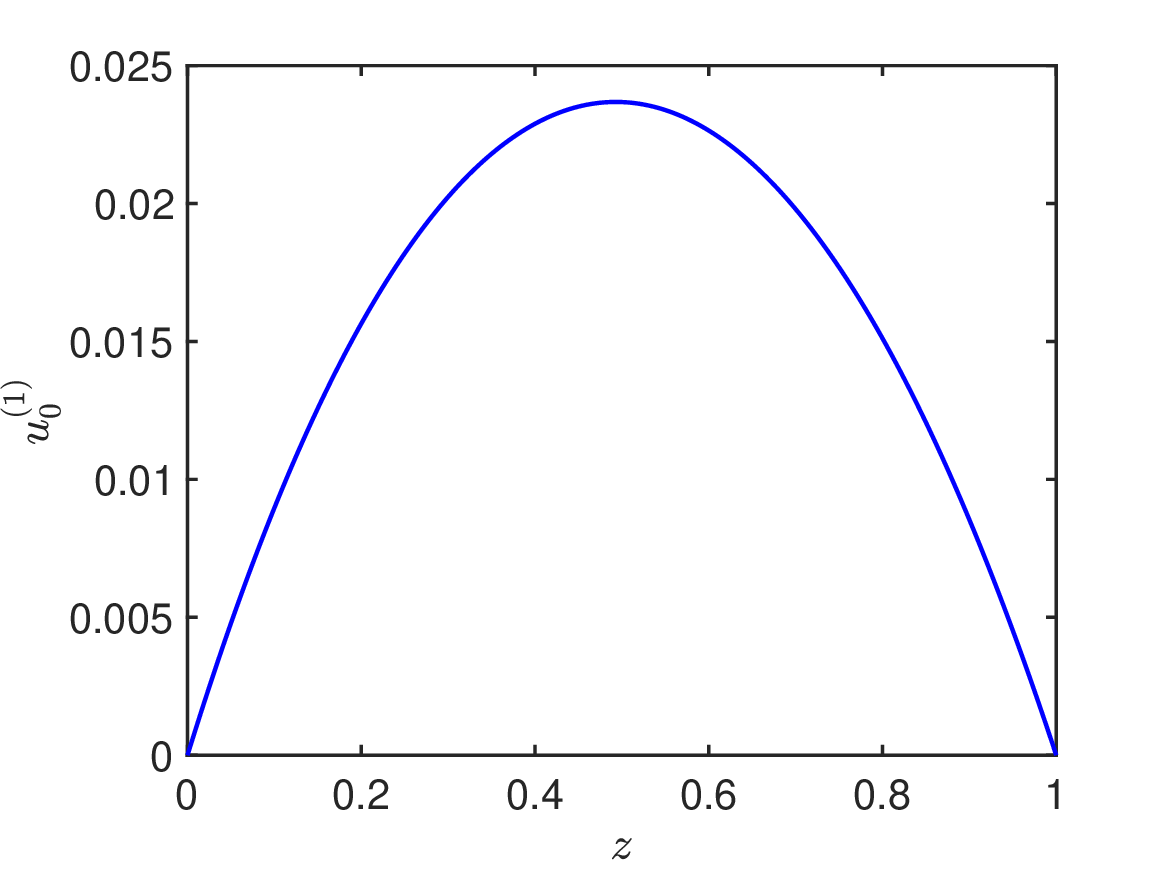}
				\put(3,72){(b)}
			\end{overpic}
		\end{minipage}
	}\\
	\subfigure {\
		\begin{minipage}[b]{.46\linewidth}
			\centering
			\begin{overpic}[scale=0.44]{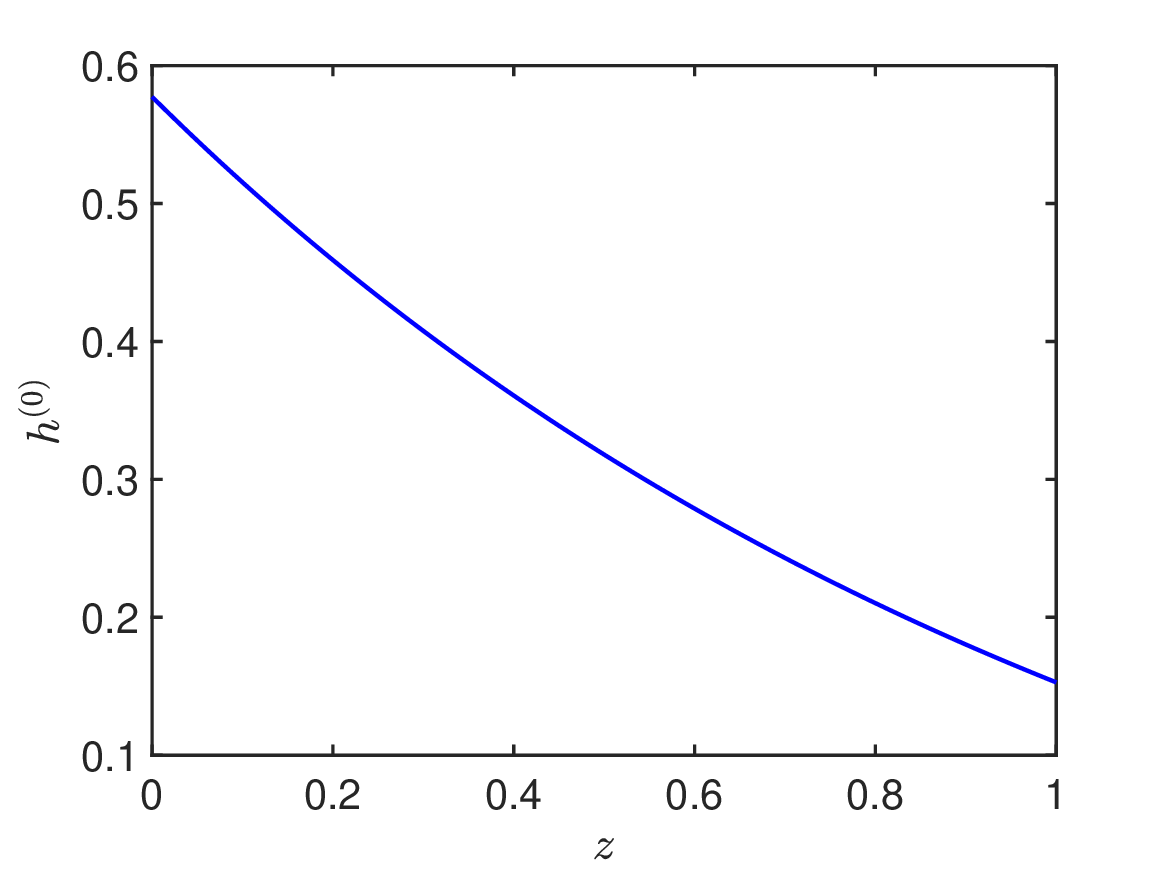}
				\put(3,72){(c)}
			\end{overpic}
		\end{minipage}
	}
	\subfigure {\
		\begin{minipage}[b]{.46\linewidth}
			\centering
			\begin{overpic}[scale=0.44]{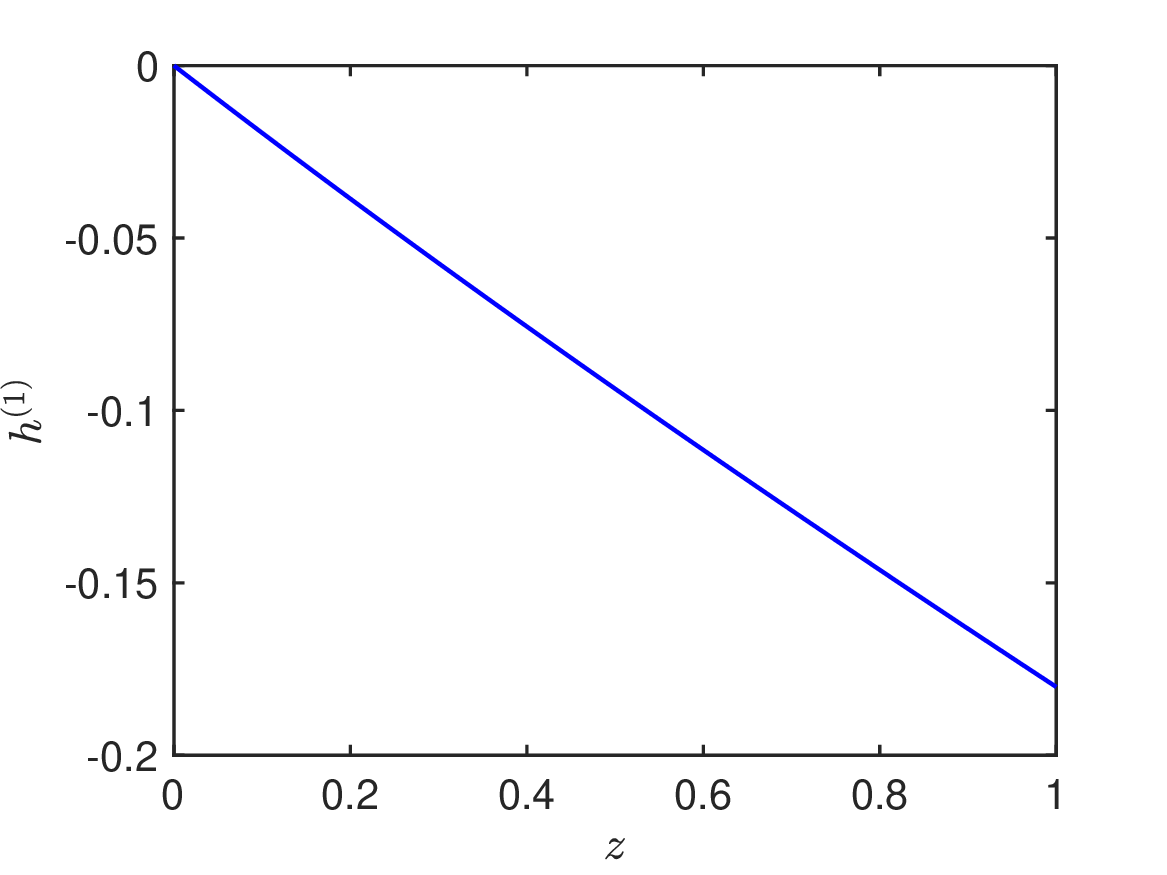}
				\put(3,72){(d)}
			\end{overpic}
		\end{minipage}
	}\\
	\subfigure {\
		\begin{minipage}[b]{.46\linewidth}
			\centering
			\begin{overpic}[scale=0.44]{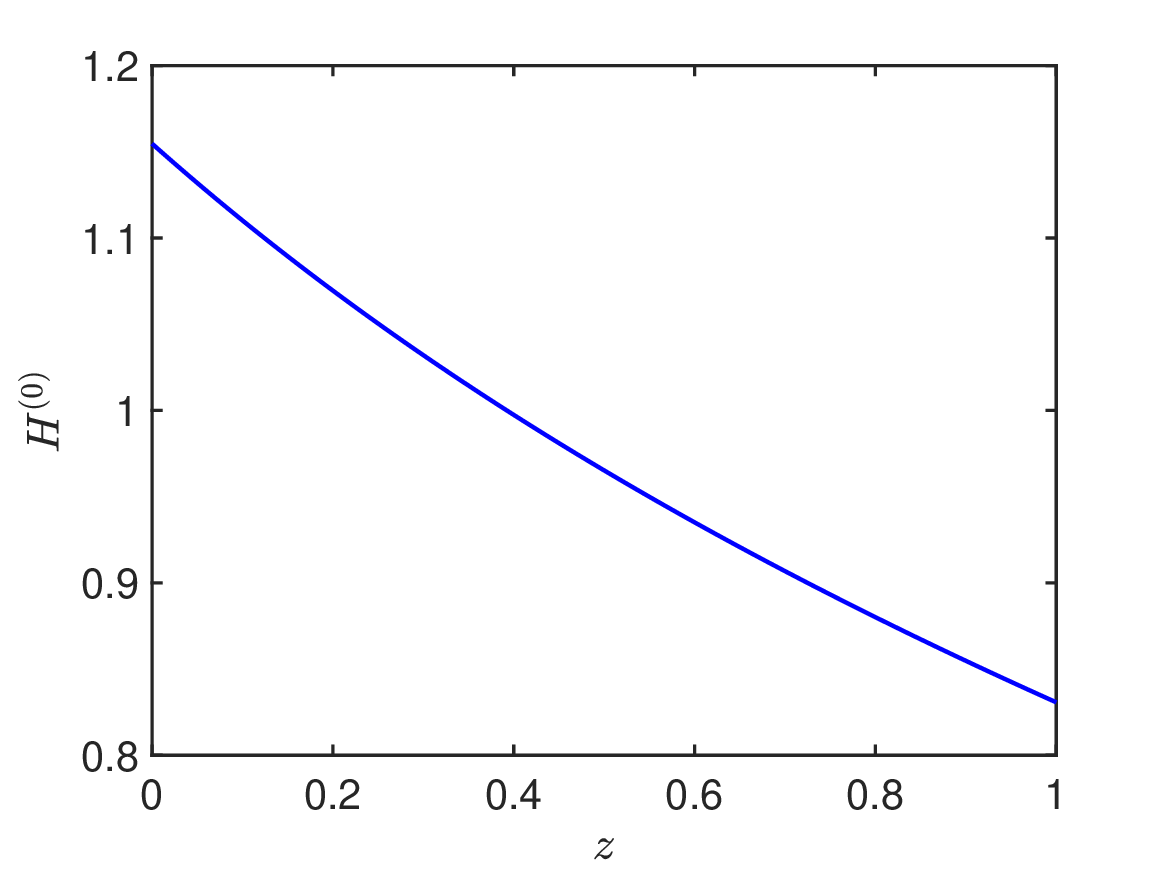}
				\put(3,72){(e)}
			\end{overpic}
		\end{minipage}
	}
	\subfigure {\
		\begin{minipage}[b]{.46\linewidth}
			\centering
			\begin{overpic}[scale=0.44]{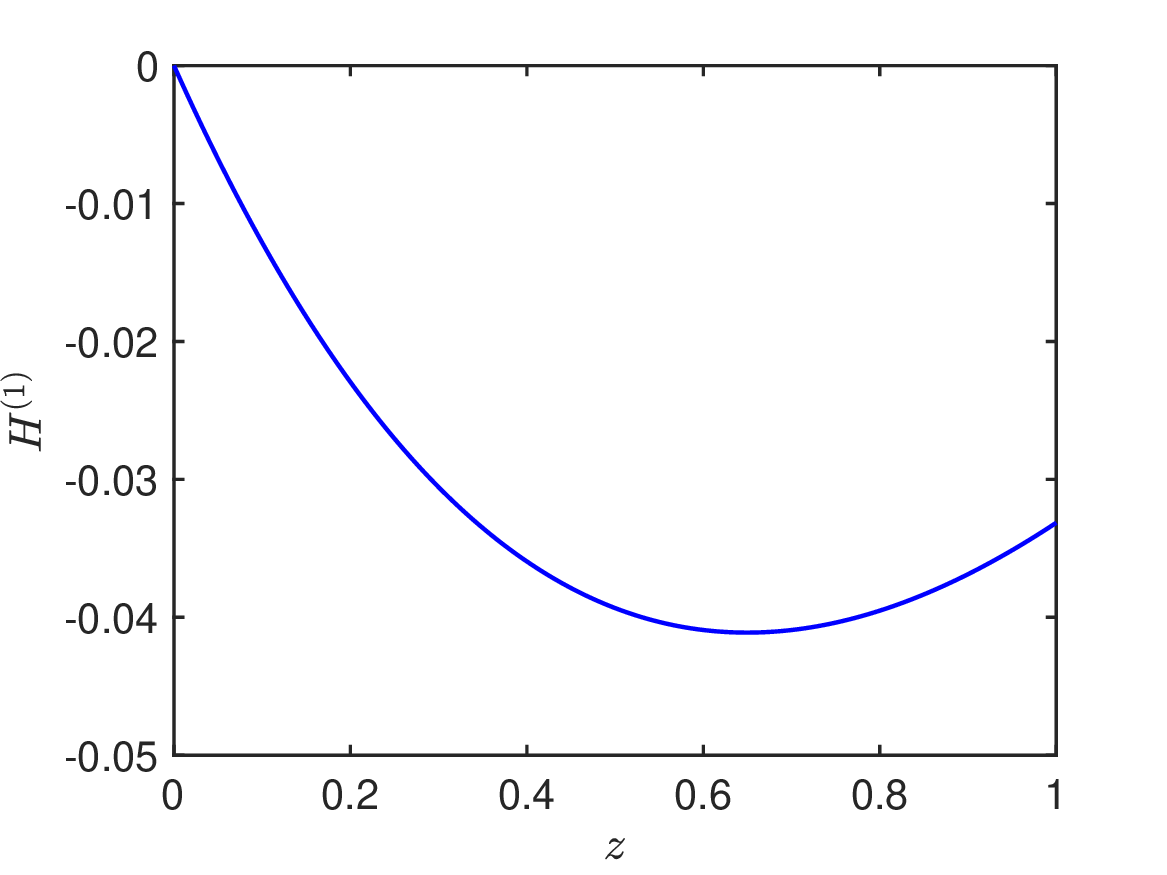}
				\put(3,72){(f)}
			\end{overpic}
		\end{minipage}
	}
	\caption{Left column: (a) the velocity $u^{(0)}_0$, (c) the hole radius $h^{(0)}$, and (e) the outer radius $H^{(0)}$ for a Newtonian fluid versus the axial distance $z$. Right column: the corrections (b) $u^{(1)}_0$, (d) $h^{(1)}$, and (f) $H^{(1)}$ to the Newtonian flow due to elastic effects. The parameters are $\mathrm{D}=1.5,\ \mathrm{Ca}=1.8,\ \mathrm{Re}=0,\ \phi=0.5,\ \alpha=0$, and $\beta=0.1$.}\label{fig:ss1}
\end{figure}
The profiles for $u^{(0)}_0,\ h^{(0)},\ H^{(0)}$ and $u^{(1)}_0,\ h^{(1)},\ H^{(1)}$ are plotted in Figure~\ref{fig:ss1} for a typical set of parameter values. Other parameters give broadly similar behaviour. The figures in the left-hand column of Figure~\ref{fig:ss1} are the leading-order velocity, hole size, and outer radius, which are the same as those for a Newtonian fluid. As a result, $u^{(0)},\ h^{(0)}$ and $H^{(0)}$ are independent of the parameters $\alpha$ and $\beta$, which characterize elasticity in the constitutive relation. The figures in the right-hand column of Figure~\ref{fig:ss1} represent the $\mathcal{O}(\mathrm{De})$ corrections to the Newtonian flow. The inlet and outlet velocities are fixed, resulting in $u^{(1)}_0$ being zero at both the inlet and outlet of the device, but $u^{(1)}_0$ is positive in the bulk meaning that elastic effects act to increase the velocity. Figure~\ref{fig:ss1}(d) shows that the elasticity acts to reduce the hole size since $h^{(1)}$ is negative. The elasticity also contributes to a reduction of the outer radius, but this is significantly weaker than the effect on the hole size. It is worth noting that the hole size at the outlet is one of the most important factors in the production of holey fibres, since the geometry at the outlet represents the output of the drawing device. As we described in the introduction, the optical properties of fibres can depend very sensitively on the output geometry. Therefore, the most important quantities from a manufacturing perspective are $h$ and $H$ at the outlet $z=1$. Furthermore, we note from \eqref{2.27a} and \eqref{2.39}$-$\eqref{2.40} that
\begin{align}
(H^2-h^2)u_0=1,\label{3.29a}
\end{align}
for the steady state. Expanding \eqref{3.29a} with respect to $\mathrm{De}$ and equating the terms of first order in $\mathrm{De}$, we obtain
\begin{align}
2(H^{(0)}H^{(1)}-h^{(0)}h^{(1)})u^{(0)}_0+(H^{(0)2}-h^{(0)2})u^{(1)}_0=0.\label{3.30a}
\end{align}
Evaluating \eqref{3.30a} at $z=1$, we obtain
\begin{align}
\frac{H^{(1)}}{h^{(1)}}=\frac{h^{(0)}}{H^{(0)}}\quad \text{at}\quad z=1.
\end{align}
Thus, we can determine $H^{(1)}$ at the outlet once $h^{(1)}$ is known. Therefore, we will focus on $h^{(1)}(z=1)$ that represents the effect that elasticity has on the hole size at the outlet in the subsequent analysis.
\begin{figure}
	\centering
	\includegraphics[width=0.6\linewidth]{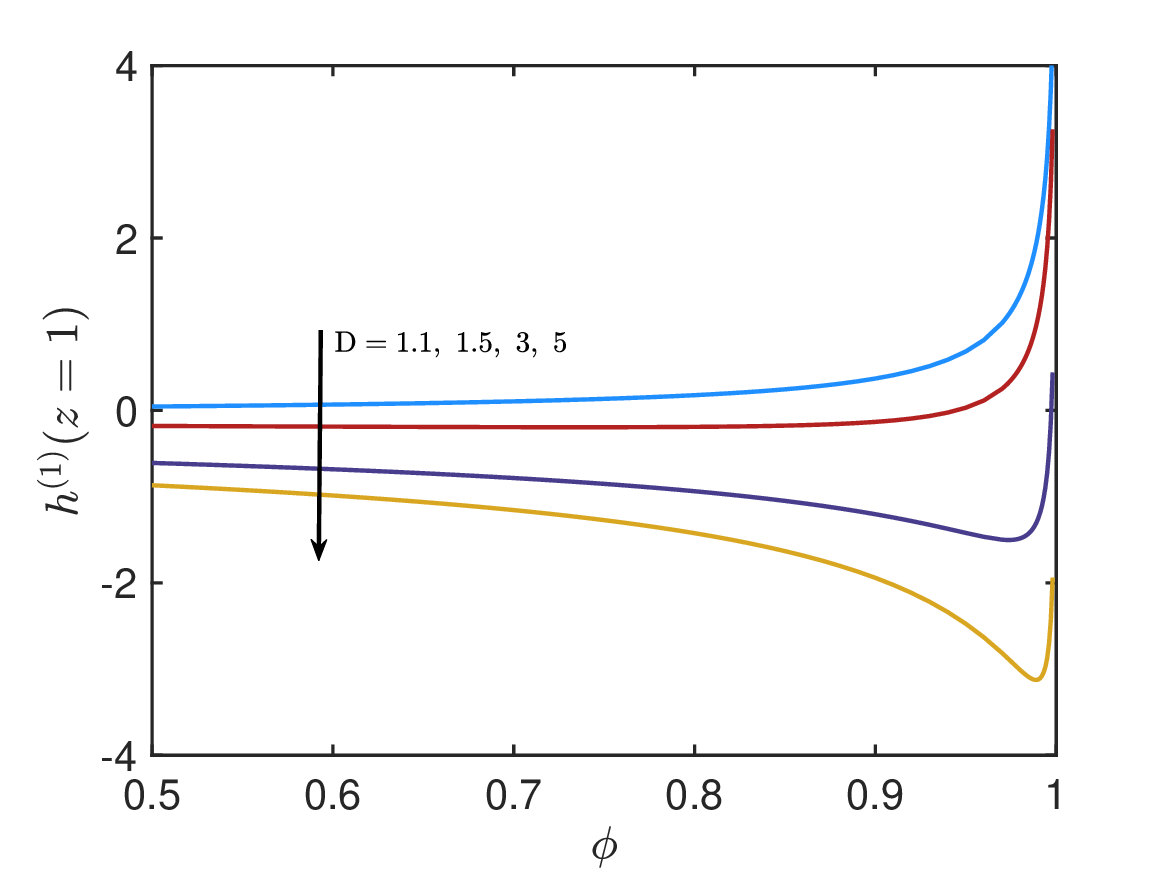}
	\caption{The effect of elasticity on the hole size at the outlet, $h^{(1)}(z=1)$, versus the ratio of inner to outer radius at the inlet, $\phi$, for different values of the draw ratio $\mathrm{D}$. Other parameters are $\mathrm{Ca}=1.8,\ \mathrm{Re}=0,\ \alpha=0,$ and $\beta=0.1.$}\label{fig:phidiffDr}
\end{figure}
\begin{figure}
	\centering
	\includegraphics[width=0.6\linewidth]{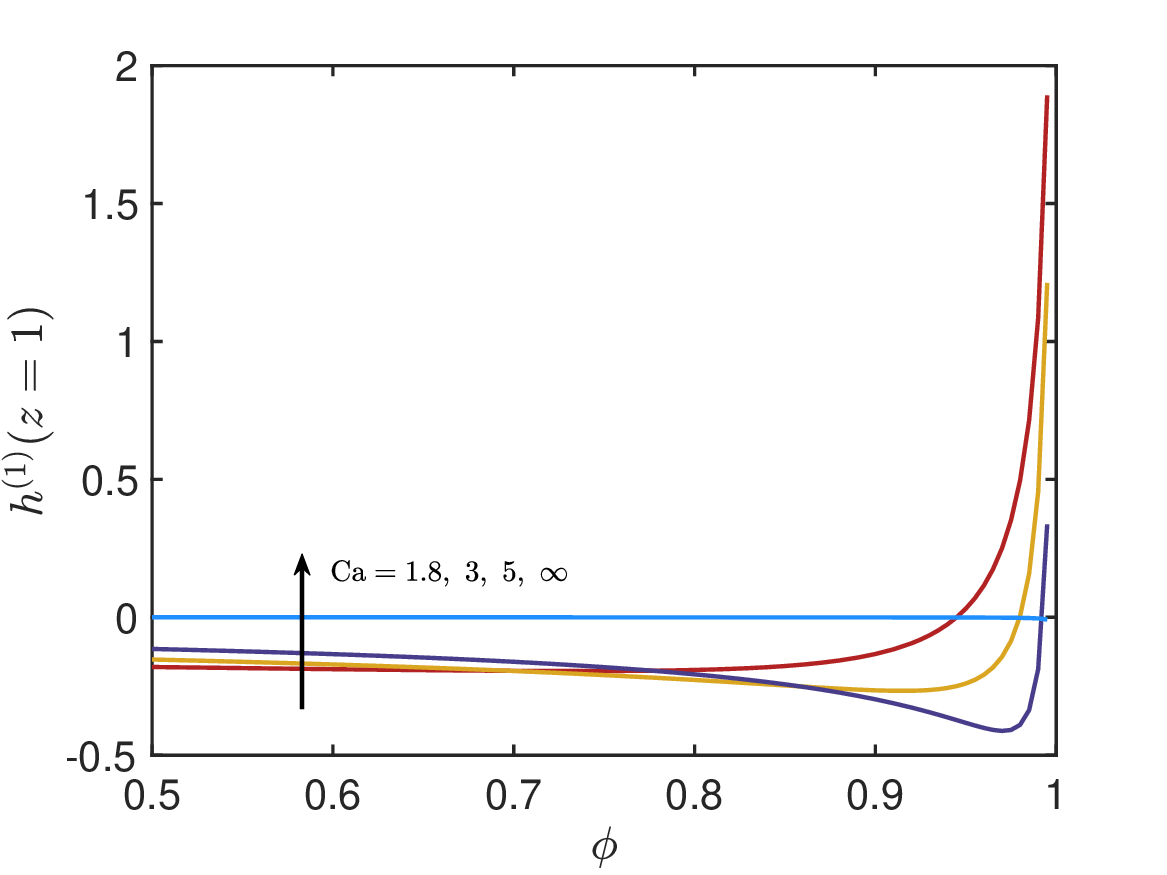}
	\caption{The effect of elasticity on the hole size at the outlet, $h^{(1)}(z=1)$, versus the ratio of inner to outer radius at the inlet, $\phi$, for different values of the Capillary number $\mathrm{Ca}$. Other parameters are $\mathrm{D}=1.5,\ \mathrm{Re}=0,\ \alpha=0,$ and $\beta=0.1.$}\label{fig:phidiffCa}
\end{figure}
In Figure~\ref{fig:phidiffDr} and Figure~\ref{fig:phidiffCa} we explore how the outlet hole size $h^{(1)}(z=1)$ depends on the inlet hole size by plotting $h^{(1)}(z=1)$ as a function of $\phi$, recalling that $\phi$ represents the ratio of inner radius to outer radius at the inlet. Note that these figures are for sufficiently large $\phi$ ($\leq0.5$) such that hole closure ($h^{(0)}=0$) does not occur. Hole closure can occur for small $\phi$, but in fibre drawing this is, perhaps with a very few exceptions, always undesirable, since if a solid thread is desired it can be more easily obtained by using a solid thread at the inlet. Therefore, in this study, we do not consider hole closure and only focus on the dynamics for sufficiently large $\phi$ such that hole closure does not occur. We plot the curves for different $\mathrm{D}$ and $\mathrm{Ca}$ in Figure~\ref{fig:phidiffDr} and \ref{fig:phidiffCa}. Figure~\ref{fig:phidiffDr} shows that an increase in $\mathrm{D}$ enhances the hole closure at the outlet due to elasticity. For very weak drawing corresponding to $\mathrm{D}=1.1$, $h^{(1)}(z=1)$ is always positive and becomes more positive as $\phi$ increases, indicating that the elasticity always plays a role in increasing the hole size. However, for moderate drawing, with $\mathrm{D}=1.5,\ 3$, and $5$, we observe that $h^{(1)}(z=1)$ is initially negative and decreases with $\phi$, indicating that the effect of elasticity in decreasing the hole size at the outlet becomes stronger as $\phi$ increases. However, for very large $\phi$, we see the opposite. This phenomenon can be also observed in the red curve in Figure~\ref{fig:phidiffCa} for $\mathrm{Ca}=1.8$ and $\mathrm{D}=1.5$. Therefore, roughly speaking, there are two different types of behaviour that can be observed in the plots: one where $\phi$ is close to $1$, which we will refer to as the ``large hole size'' case, and another where $\phi$ is not close to $1$, which we will refer to as the ``moderate hole size'' case. The subsequent sections will concentrate on examining these two cases. Furthermore, Figure~\ref{fig:phidiffCa} shows that for sufficiently small values of $\phi$ an increase in $\mathrm{Ca}$ will decrease the hole closure at the outlet due to elasticity. However, the opposite trend is observed for sufficiently large values of $\phi$. In addition, as $\mathrm{Ca}\to\infty$, \eqref{3.9} and \eqref{3.26} imply that $C^{(0)}$ and $C^{(1)}$ will tend to zero, and one can integrate the steady version of \eqref{3.28} from $z=0$ to $z=1$ to see that $h^{(1)}(z=1)=0$. Hence, the hole size at the outlet for viscoelastic fluids will tend to that for a Newtonian fluid and hence $h^{(1)}(z=1)$ will tend to zero as $\mathrm{Ca}\to\infty$, as shown in Figure~\ref{fig:phidiffCa}.

\subsubsection{Moderate and large hole sizes}
\begin{figure}
	\centering
	\includegraphics[width=0.6\linewidth]{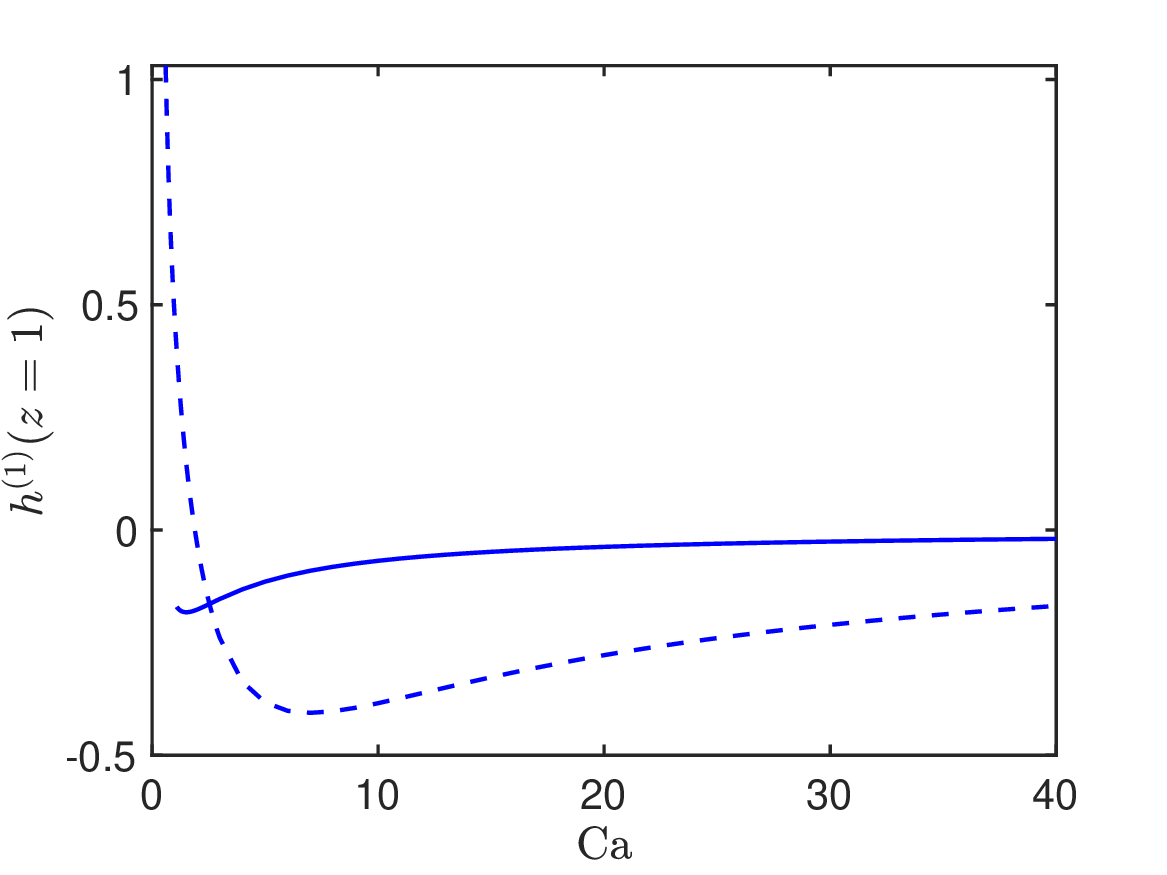}
	\caption{The effect of elasticity on the hole size at the outlet versus capillary number $\mathrm{Ca}$. The solid curve is for $\phi=0.5$, the dashed curve is for $\phi=0.95$. Other parameters are $\mathrm{D}=1.5,\ \mathrm{Re}=0,\ \alpha=0,$ and $\beta=0.1$.}\label{fig:Cadiff}
\end{figure}

\begin{figure}
	\centering
	\includegraphics[width=0.6\linewidth]{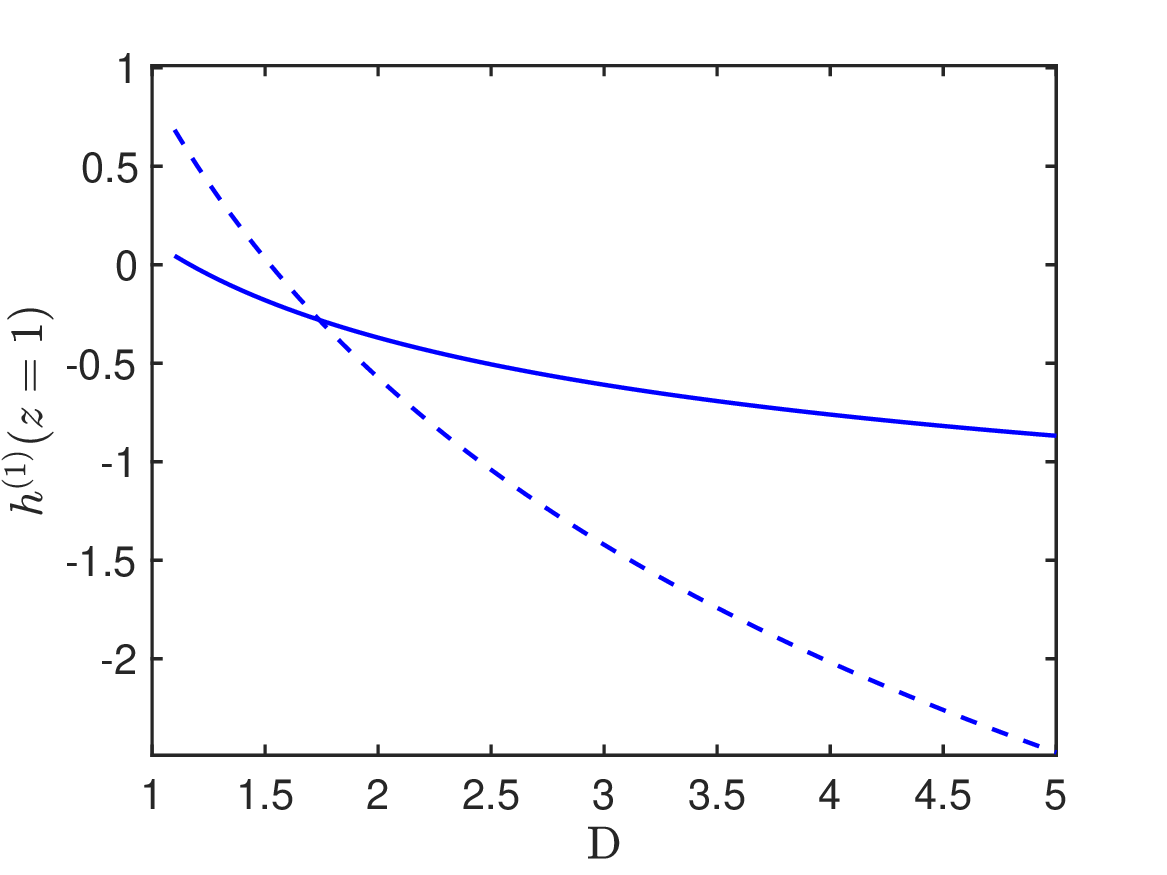}
	\caption{The effect of elasticity on the hole size at the outlet versus draw ratio $\mathrm{D}$. The solid curve is for $\phi=0.5$, the dashed curve is for $\phi=0.95$. Other parameters are $\mathrm{Re}=0,\ \mathrm{Ca}=1.8,\ \alpha=0,$ and $\beta=0.1.$}\label{fig:Drdiff}
\end{figure}

\begin{figure}
	\centering
	\includegraphics[width=0.6\linewidth]{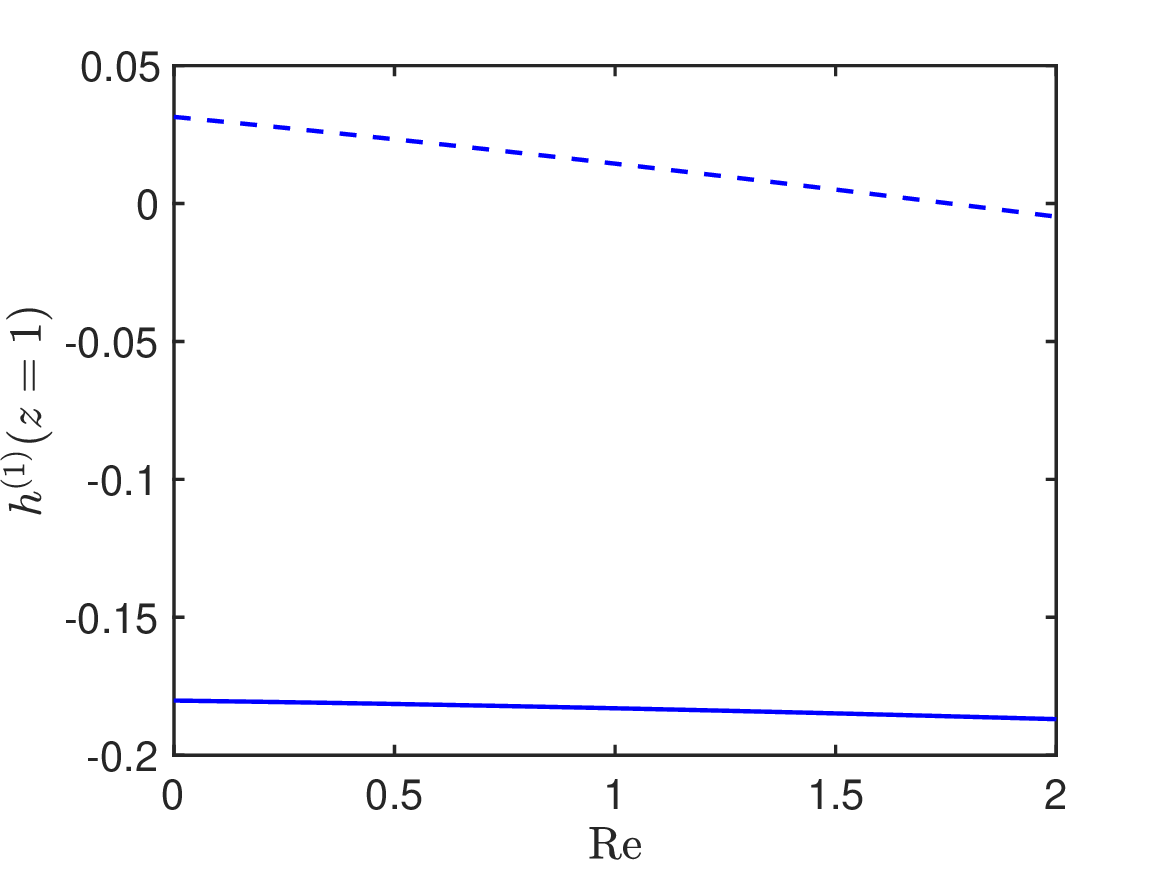}
	\caption{The effect of elasticity on the hole size at the outlet versus Reynolds number $\mathrm{Re}$. The solid curve is for $\phi=0.5$, the dashed curve is for $\phi=0.95$. Other parameters are $\mathrm{D}=1.5,\ \mathrm{Ca}=1.8,\ \alpha=0,$ and $\beta=0.1.$}\label{fig:Rediff}
\end{figure}

\begin{figure}
	\centering
	\includegraphics[width=0.6\linewidth]{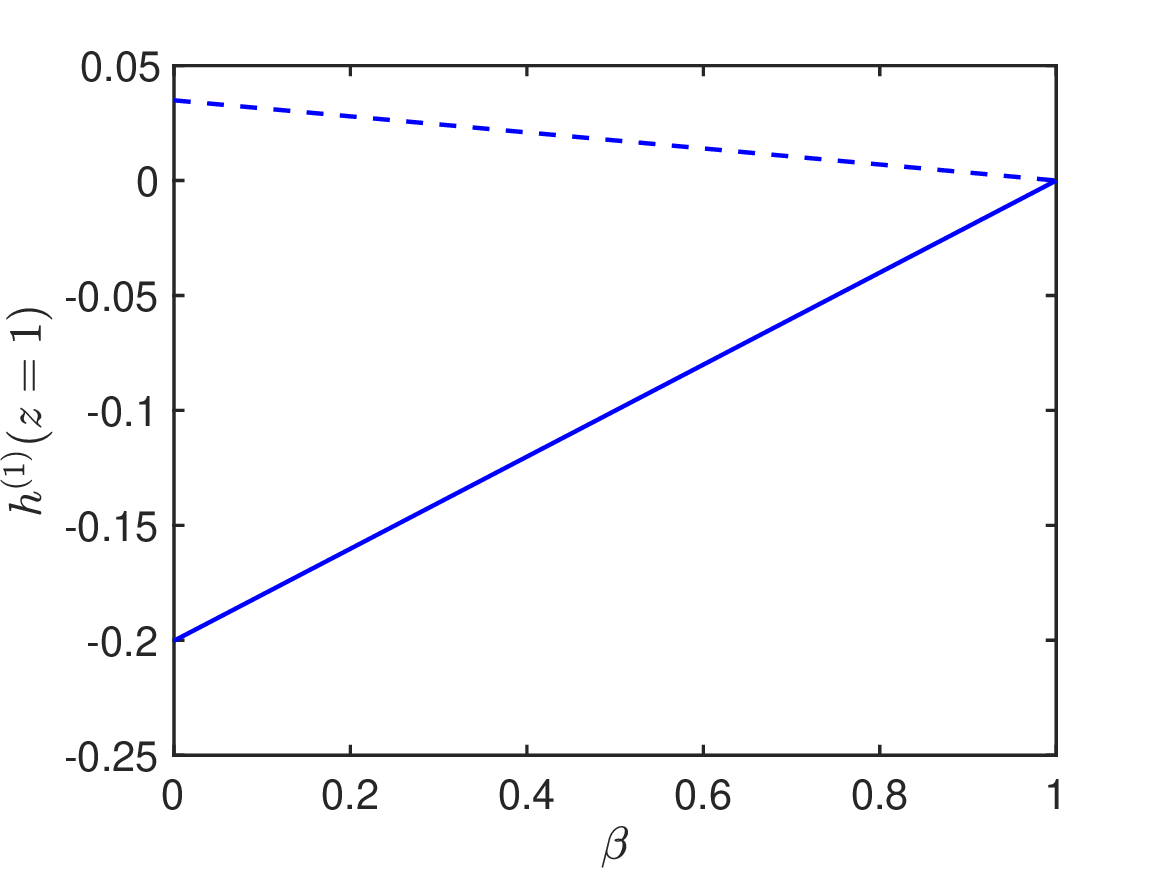}
	\caption{The effect of elasticity on the hole size at the outlet versus solvent to bulk viscosity $\beta$. The solid curve is for $\phi=0.5$, the dashed curve is for $\phi=0.95$. Other parameters are $\mathrm{D}=1.5,\ \mathrm{Ca}=1.8,\ \mathrm{Re}=0,$ and $\alpha=0.$}\label{fig:betadiff}
\end{figure}

\begin{figure}
	\centering
	\includegraphics[width=0.6\linewidth]{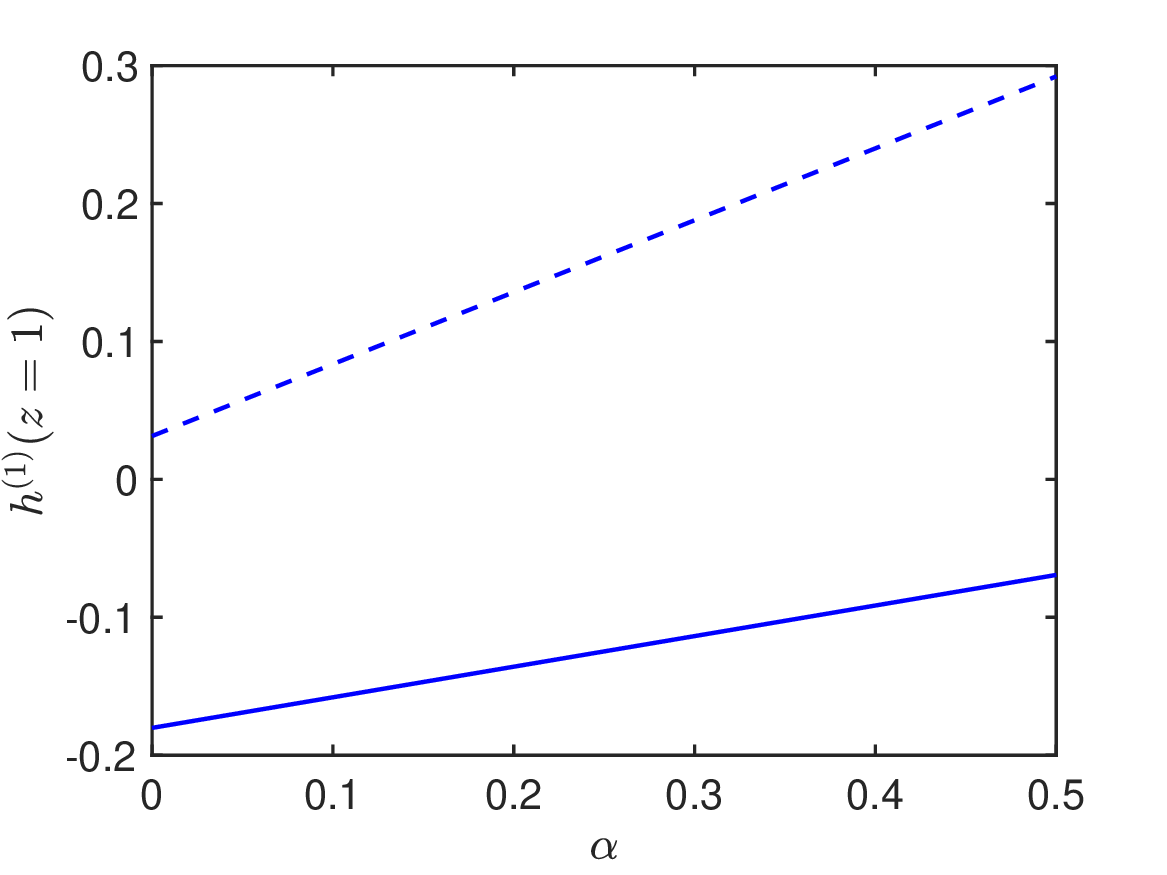}
	\caption{The effect of elasticity on the hole size at the outlet versus mobility factor $\alpha$. The solid curve is for $\phi=0.5$, the dashed curve is for $\phi=0.95$. Other parameters are $\mathrm{D}=1.5,\ \mathrm{Ca}=1.8,\ \mathrm{Re}=0,$ and $\beta=0.1.$}\label{fig:alphadiff}
\end{figure}
In the Figures~\ref{fig:Cadiff}$-$\ref{fig:alphadiff}, we compare the hole closure at the outlet due to elasticity for moderate ($\phi=0.5$) and large ($\phi=0.95$) inlet hole sizes. In particular, we examine how the various parameters $\mathrm{Ca},\ \mathrm{D},\ \mathrm{Re},\ \beta,$ and $\alpha$ affect $h^{(1)}(z=1)$. In Figure~\ref{fig:Cadiff}, we examine the effect of capillary number ($\mathrm{Ca}$). Here, we start with $\mathrm{Ca}\approx1.1$ for the case of a moderate hole size and $\mathrm{Ca}\approx0.6$ for the case of a large hole size, respectively, as the hole will close for smaller capillary numbers corresponding to very strong surface tension. For the moderate value of $\phi$ (solid curve), $h^{(1)}(z=1)$ is always negative indicating that the hole size at the outlet is consistently smaller than for a pure Newtonian fluid. For large $\phi$ (dashed curve) and sufficiently small $\mathrm{Ca}$, $h^{(1)}(z=1)$ is positive indicating that the hole size at the outlet is bigger than the Newtonian one. Furthermore, we note that $h^{(1)}(z=1)$ is not a monotonic function of $\mathrm{Ca}$. For sufficiently small $\mathrm{Ca}$, $h^{(1)}(z=1)$ reduces as $\mathrm{Ca}$ increases. However, this trend reverses beyond a critical value of $\mathrm{Ca}$. This non-monotonic behaviour can also be observed in Figure~\ref{fig:phidiffCa}. As shown in Figure~\ref{fig:phidiffCa}, $h^{(1)}(z=1)\to0$ when $Ca\to\infty$ (negligible surface tension) for all $\phi$.

Figure~\ref{fig:Drdiff} shows the effect of draw ratio $\mathrm{D}$ and that there is a critical value of $\mathrm{D}$ above which elastic effects decrease and below which elastic effects increase the hole size at the outlet. For the moderate value of $\phi$ (solid curve) elastic effects only act to very weakly increase the hole size for values of $\mathrm{D}$ quite close to $1$. Whereas, for the large value of $\phi$ (dashed curve) the effect is much stronger and acts over a larger range of $\mathrm{D}$ values. An increase in the draw ratio uniformly enhances the hole closure at the outlet due to elasticity. This phenomenon can also be observed in Figure~\ref{fig:phidiffDr}. 

The effect of inertia is shown in Figure~\ref{fig:Rediff}. For $\phi=0.5$ (solid curve), we observe that $h^{(1)}(z=1)$ is always negative indicating that elastic effects enhance hole closure at the outlet. On the other hand, for $\phi=0.95$ (dashed curve) we observe that $h^{(1)}(z=1)$ is positive for $\mathrm{Re}\lesssim1.8$, indicating that elastic effects suppress hole closure. However, for $\mathrm{Re}\gtrsim1.8$ we see that the opposite is true. Nevertheless, as $\mathrm{Re}$ varies the effect on $h^{(1)}(z=1)$ is relatively weak.

Figure~\ref{fig:betadiff} shows how $\beta$ (solvent to bulk viscosity ratio) affects $h^{(1)}(z=1)$. We note that the maximum value $\beta$ can take is $\beta=1$, and this corresponds to a purely Newtonian fluid (only solvent). Therefore, as $\beta$ tends to unity, the elastic effects will tend to zero and we will get $h^{(1)}(z=1)\to0$. For $\phi=0.5$, $h^{(1)}(z=1)$ is negative, indicating that the elasticity acts to enhance hole closure. As $\beta$ increases, the effect of elasticity becomes weaker. While for $\phi=0.95$, $h^{(1)}(z=1)$ is positive, indicating that the elasticity suppresses hole closure. As $\beta$ increases towards unity the effects of elasticity on the hole closure monotonically decrease to zero.

In Figure~\ref{fig:alphadiff} we show how $\alpha$ (mobility factor) affects $h^{(1)}(z=1)$. We note that in the Giesekus model, physical constraints demand that $\alpha<0.5$ \citep{morozov2015introduction}. For $\phi=0.5$ elastic effects enhance hole closure, but this becomes weaker as $\alpha$ increases. On the other hand, elastic effects suppress hole closure for $\phi=0.95$ and this suppression becomes stronger as $\alpha$ increases. We will return to this phenomenon in the next subsection.

\subsection{Assessing the effects of different contributions to the hole size at the outlet}
It is well known that elastic effects suppress the surface-tension-driven pinching that occurs for threads without holes \citep{entov1997effect,li2003drop}. Therefore, one might guess that elastic effects will suppress the surface-tension-driven hole closure that occurs for threads with holes. However, from Figure~\ref{fig:phidiffDr} and \ref{fig:phidiffCa}, we surprisingly found that elastic effects enhance hole closure, provided $\phi$ is not too large and $\mathrm{D}$ is not close to unity. On the other hand, for threads with large initial hole size or draw ratio $\mathrm{D}$ close to unity, the elasticity has the opposite effect. To provide further insight into this behaviour, we elucidate the relative importance of different contributions to $h^{(1)}(z=1)$.   We remind the reader that the radial flow is of the form $v_0=-r u_{0,z}/2+C/r$, where $C$ represents the strength of the radial flow induced by surface tension forces, which we expanded as an asymptotic series in terms of the Deborah number, $C=C^{(0)}+\mathrm{De}\ C^{(1)}$. Here, $C^{(0)}$ represents the strength of the radial flow induced by surface tension for zero Deborah number and $C^{(1)}$ represents the correction to this radial flow that arises from elastic effects. The inner surface must necessarily have larger curvature than the outer surface, and this will generate an inward flow meaning that $C^{(0)}<0$. 

At leading order, the inward surface tension forces generate a flow that is opposed by viscous stresses. By considering the leading-order $r-$momentum equation \eqref{2.30} and normal dynamic boundary conditions \eqref{2.34}$-$\eqref{2.35} we obtain
\begin{align}
\int_{h^{(0)}}^{H^{(0)}}\frac{N^{(0)}_2}{r}dr=\frac{1}{Ca}(\frac{1}{h^{(0)}}+\frac{1}{H^{(0)}}),\label{3.32a}
\end{align}
where $N^{(0)}_2=\sigma^{(0)}_{rr0}-\sigma^{(0)}_{\theta\theta0}$ is the second normal stress difference to leading order in $\mathrm{De}$, with $\sigma^{(0)}_{rr0}=-p+2\beta v^{(0)}_{0,r}+\tau^{(0)}_{rr0}$ and $\sigma^{(0)}_{\theta\theta0}=-p+2\beta v^{(0)}_0/r+\tau^{(0)}_{\theta\theta0}$. Equation \eqref{3.32a} shows that, at leading order, the surface tension force is balanced purely by the second normal stress difference.

By setting $\partial_t\equiv 0$ and integrating \eqref{3.28} with respect to $z$, we obtain
\begin{align}
h^{(1)}(z=1)=\frac{1}{h^{(0)}(z=1)\mathrm{D}} \int_{0}^{1} C^{(1)}dz.\label{4.14}
\end{align}
This shows that $h^{(1)}(z=1)$ is directly determined by the elastic correction to the surface-tension-driven radial flow. In some sense \eqref{4.14} simply reflects the fact that an increased inward radial flow due to elastic effects will combine with the kinematic condition \eqref{2.10} to enhance hole closure.

In order to understand the physical mechanism that is associated with $C^{(1)}$, we subtract the two first-order normal boundary conditions \eqref{3.20} and \eqref{3.21}, then using \eqref{3.3}, \eqref{3.19}, and integrating by parts, we obtain
\begin{align}
C^{(1)}=\frac{h^{(0)2} H^{(0)2}}{2\beta (H^{(0)2}-h^{(0)2})}\int_{h^{(0)}}^{H^{(0)}}\frac{T^{(1)}_2}{r}dr+\frac{h^{(0)2} H^{(1)}-H^{(0)2}h^{(1)}}{2 \beta \mathrm{Ca}(H^{(0)}-h^{(0)})^2},\label{3.22}
\end{align}
where we define $T_2=\tau_{rr0}-\tau_{\theta\theta0}$ to be the elastic component of the second normal stress difference and expand $T_2$ in the form $T_2=T_2^{(0)}+\mathrm{De} T_2^{(1)}$. This equation shows that the elastic correction to $C^{(1)}$ and hence the hole size at the outlet only depend on the stress via $T^{(1)}_2$. Second normal stress differences are often ignored in viscoelastic flows, because first normal stress differences are generally more important. Nevertheless, \cite{maklad2021review} give a very complete description of their importance and the specific physical role they can play. As we can see in \eqref{3.22} the effect of elasticity on hole evolution is directly related to the second normal stress difference and is independent of the first normal stress difference.

We note that \eqref{3.22} contains two terms on the right-hand side. The first term involves the elastic component of the second normal stress difference, while the second term comes from the linearized surface tension terms in \eqref{3.20} and \eqref{3.21} that arise from expanding $h$ and $H$ as a series in $\mathrm{De}$. This second term does not directly involve the elastic stresses and is therefore simply a response to the elastic forces rather than a driving mechanism for the evolution of the hole.

Equations \eqref{3.15}$-$\eqref{3.16} show how the first-order corrections to the elastic component of the second normal stress difference, given by $T^{(1)}_2$, arise as an elastic response to the stresses generated by $\tau^{(0)}_{rr0}$ and $\tau^{(0)}_{\theta\theta0}$ for the zero Deborah number flow. We can then use \eqref{3.4}$-$\eqref{3.5} and \eqref{3.15}$-$\eqref{3.16} to rewrite $C^{(1)}$ in the form:
\begin{align}
C^{(1)}&=\underbrace{(1-\beta)u^{(0)}_0 C^{(0)}_{,z}}_{\textcircled{\small 1}}+\underbrace{3(1-\beta)C^{(0)}u^{(0)}_{0,z}+(-1)(1-\beta)\frac{(h^{(0)2}+H^{(0)2})C^{(0)2}}{h^{(0)2}H^{(0)2}}}_{\textcircled{\small 2}}\nonumber\\
&+\underbrace{2\alpha(-1)(1-\beta)C^{(0)}u^{(0)}_{0,z}}_{\textcircled{\small 3}}+\underbrace{\Bigg[\frac{h^{(0)2}H^{(1)}-H^{(0)2}h^{(1)}}{2\mathrm{Ca}(H^{(0)}-h^{(0)})^2}\Bigg]}_{\textcircled{\small 4}}.\label{4.16}
\end{align}
Equation \eqref{4.16} shows that $C^{(1)}$ arises from complicated interactions between the upper-convected derivative and the full stress tensor. We have labeled different terms in \eqref{4.16} to help us understand the mechanism that drives hole evolution. Terms ${\textcircled{\small 1}},\ {\textcircled{\small 2}}$ and $\textcircled{\small 3}$ all contain $C^{(0)}$ which represents the strength of the radial flow induced by surface tension in the zero Deborah number flow limit. As explained above, $C^{(0)}$ will always be negative and represents an inward flow. In this physical description, we will only focus on how the radial flow associated with $C^{(0)}$ induces leading-order elastic stresses, and how this affects $C^{(1)}$. The radial velocity associated with axial stretching (the $-r u^{(0)}_{0,z}/2$ term in \eqref{3.2}) can be dealt with in a similar way.

In order to understand how elastic effects play a role in the evolution of the hole size we consider the constitutive relations \eqref{3.15}$-$\eqref{3.16} that represent the $\mathcal{O}(\mathrm{De})$ corrections to the stresses. These equations show that the leading-order stresses are advected by the zero Deborah number flow in both axial and radial directions and this advection causes changes to the stress. The constitutive relations indicate that these changes will be opposed by $\mathcal{O}(\mathrm{De})$ elastic stresses $\tau^{(1)}_{rr0}$, $\tau^{(1)}_{\theta\theta0}$. To understand the physical origin of the terms in \eqref{4.16} we need to examine the radial and axial advection terms that give rise to $\mathcal{O}(\mathrm{De})$ elastic stresses. We begin by considering the radial advection. In the zero Deborah number approximation, surface tension induces a radial velocity of the form $C^{(0)}/r$ (see \eqref{3.2}), with $C^{(0)}< 0$. Using \eqref{3.2}$-$\eqref{3.5} we see that the leading-order second normal stress difference $N_2^{(0)}$ that opposes this radial flow is in the positive $r-$direction and decreases as $r$ increases. Hence, the radially inward advective flow will move fluid particles from low positive stress to high positive stress and thus increases the stress. At $\mathcal{O}(\mathrm{De})$, the elastic response $T^{(1)}_2$ will oppose this stress increase and thus will be negative. Therefore, this effect means that elastic effects associated with radial advection make the hole smaller. This along with the other terms in the upper-convected derivative represents the physical origin of term $\textcircled{\small 2}$ in \eqref{4.16}.

We next consider axial advection which advects in the positive $z-$direction since $u^{(0)}_0>0$. Using \eqref{3.2}$-$\eqref{3.5} we see that the leading-order second normal stress difference is positive and proportional to $C^{(0)}$. If $C^{(0)}_{,z}>0$, then axial advection terms will move fluid particles from regions of large positive stress to regions of low positive stress, and thus reduce the stress. At $\mathcal{O}(\mathrm{De})$, elastic stresses will oppose this change in stress and so the elastic response $T^{(1)}_2$ will be positive. This means that elastic effects associated with axial advection act to make the hole larger if $C^{(0)}_{,z}>0$. This is the physical origin of term $\textcircled{\small 1}$ in \eqref{4.16}.

\begin{figure}
	\centering
	\subfigure {\
		\begin{minipage}[b]{.46\linewidth}
			\centering
			\begin{overpic}[scale=0.44]{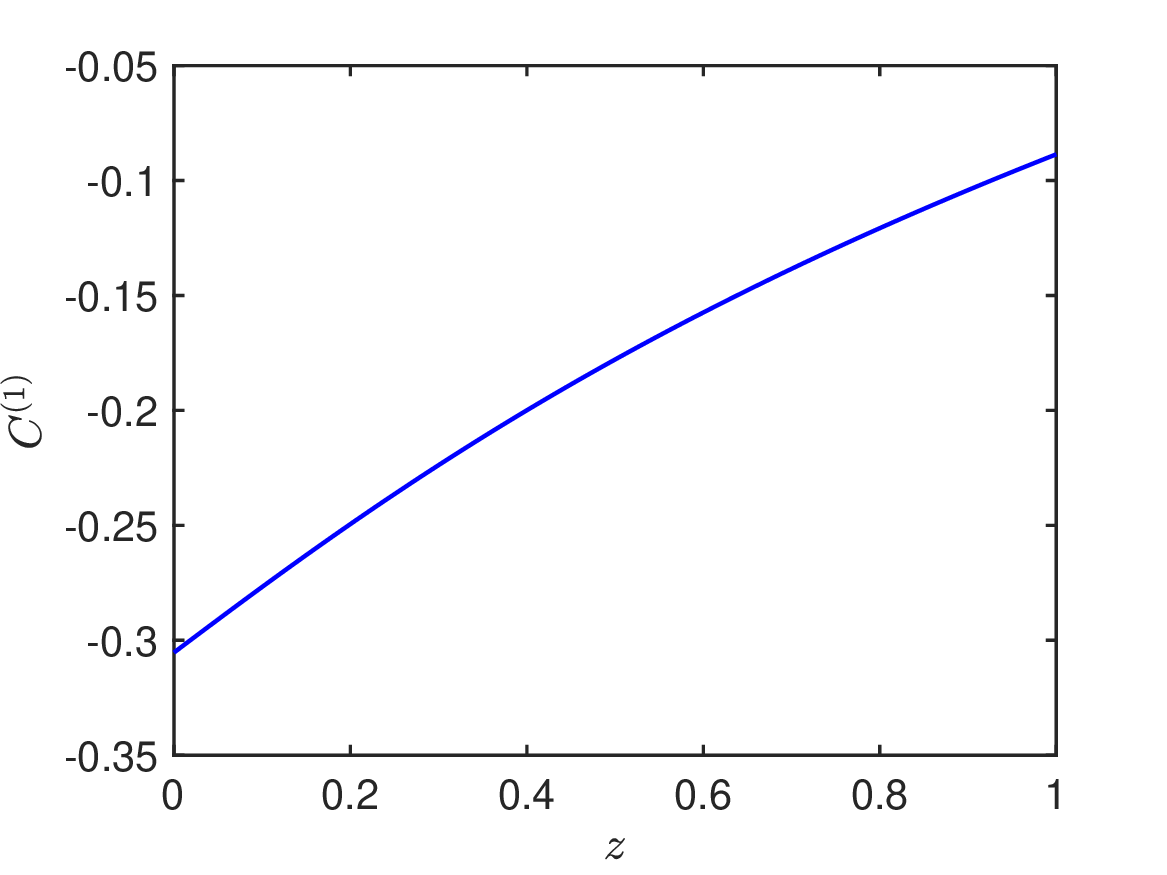}
				\put(3,72){(a)}
			\end{overpic}
		\end{minipage}
	}
	\subfigure {\
		\begin{minipage}[b]{.46\linewidth}
			\centering
			\begin{overpic}[scale=0.44]{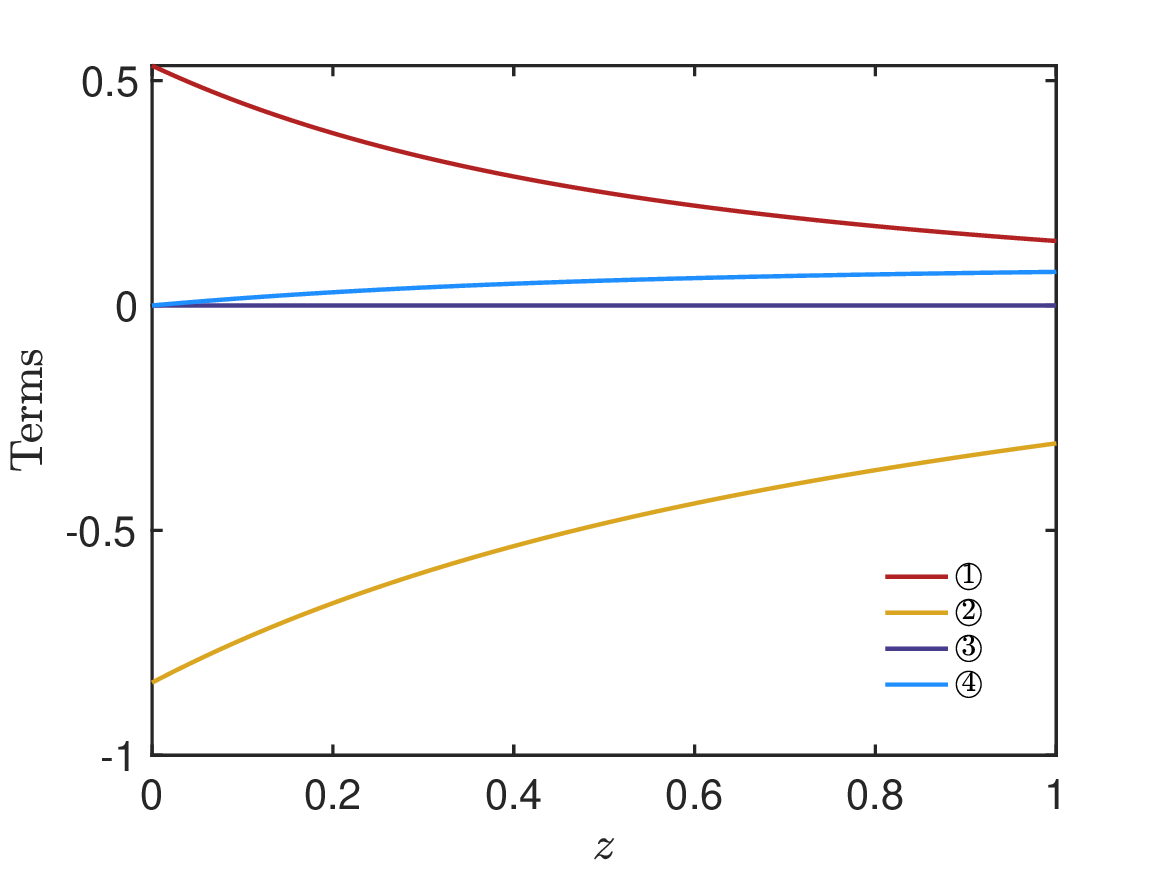}
				\put(3,72){(b)}
			\end{overpic}
		\end{minipage}
	}\\
	\subfigure {\
		\begin{minipage}[b]{.46\linewidth}
			\centering
			\begin{overpic}[scale=0.44]{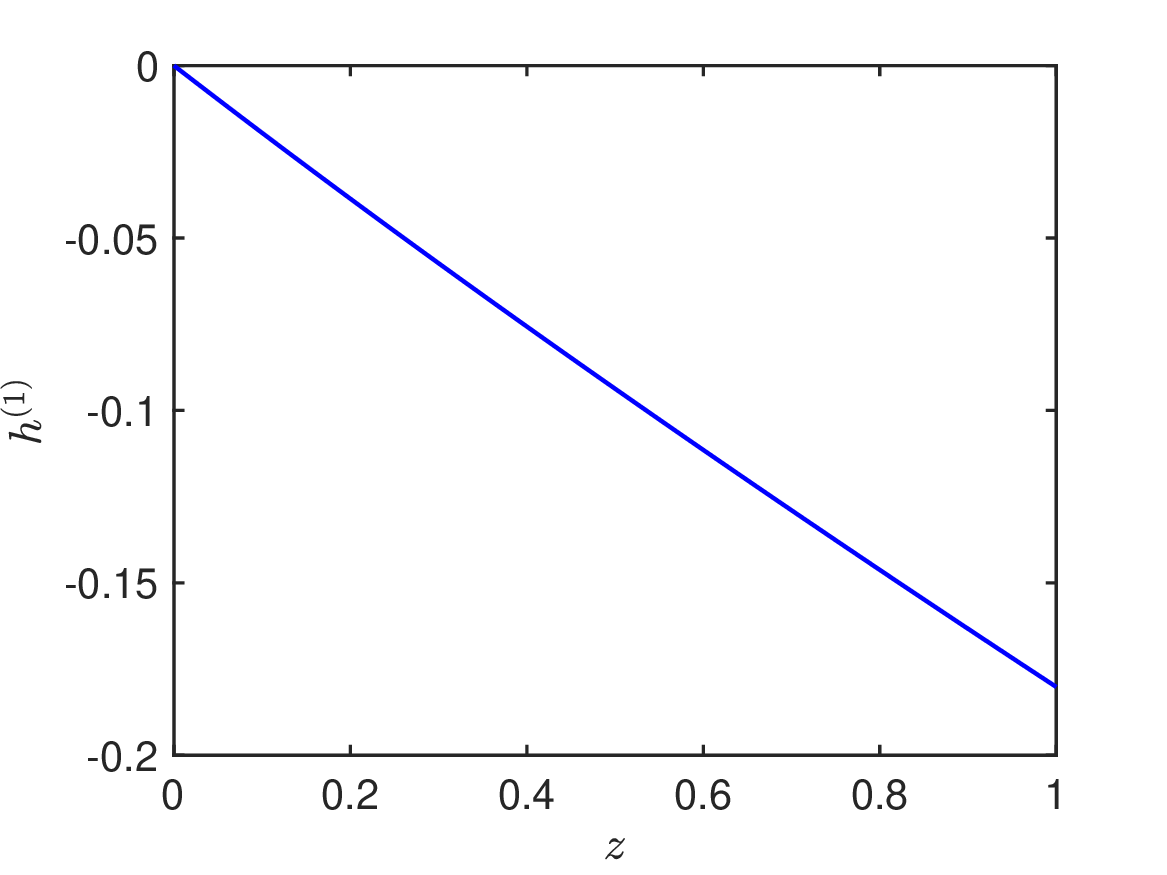}
				\put(3,72){(c)}
			\end{overpic}
		\end{minipage}
	}
	\caption{(a) The quantity $C^{(1)}$, (b) the terms $\textcircled{\small 1}-\textcircled{\small 4}$ in \eqref{4.16} , and (c) $h^{(1)}$ versus $z$ for $\mathrm{D}=1.5,\ \mathrm{Ca}=1.8,\ \phi=0.5,\ \mathrm{Re}=0,\ \alpha=0,\ \beta=0.1$.}\label{fig:ss5}
\end{figure}

\begin{figure}
	\centering
	\subfigure {\
		\begin{minipage}[b]{.46\linewidth}
			\centering
			\begin{overpic}[scale=0.44]{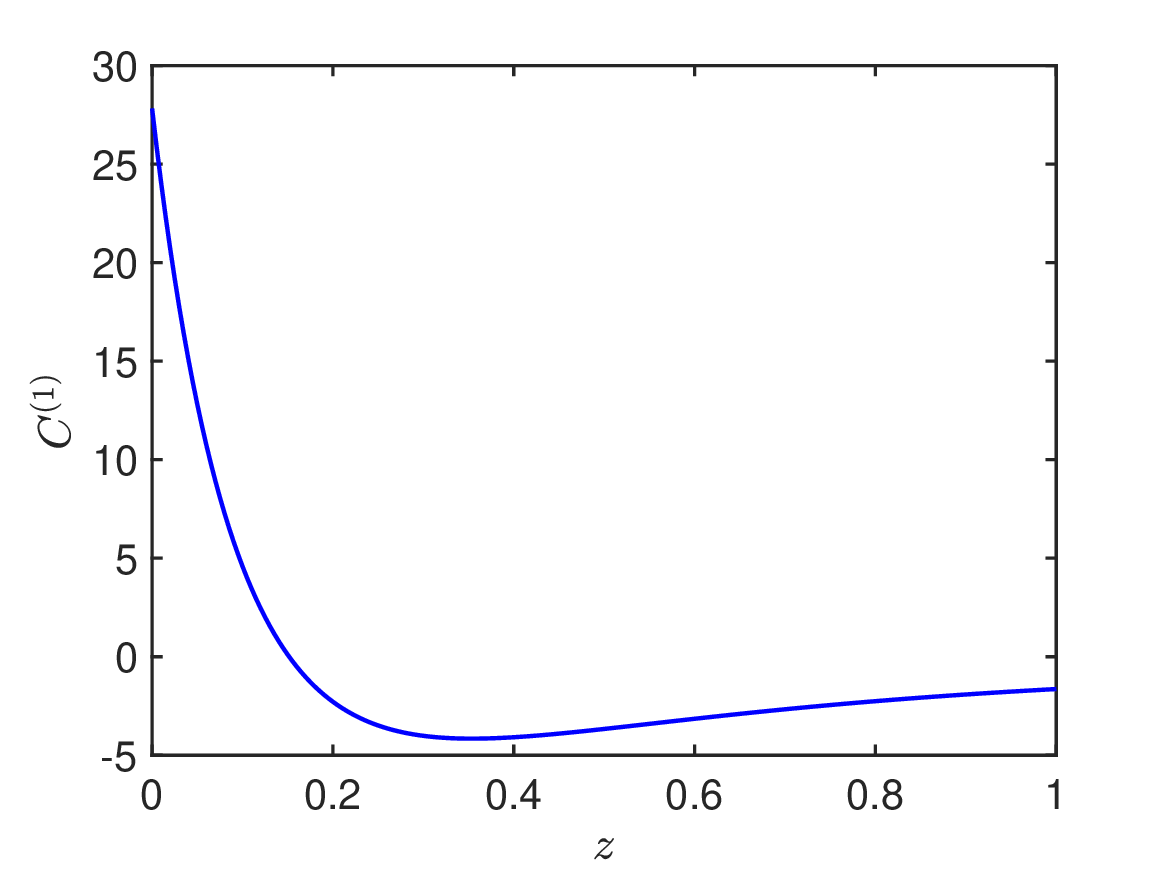}
				\put(3,72){(a)}
			\end{overpic}
		\end{minipage}
	}
	\subfigure {\
		\begin{minipage}[b]{.46\linewidth}
			\centering
			\begin{overpic}[scale=0.44]{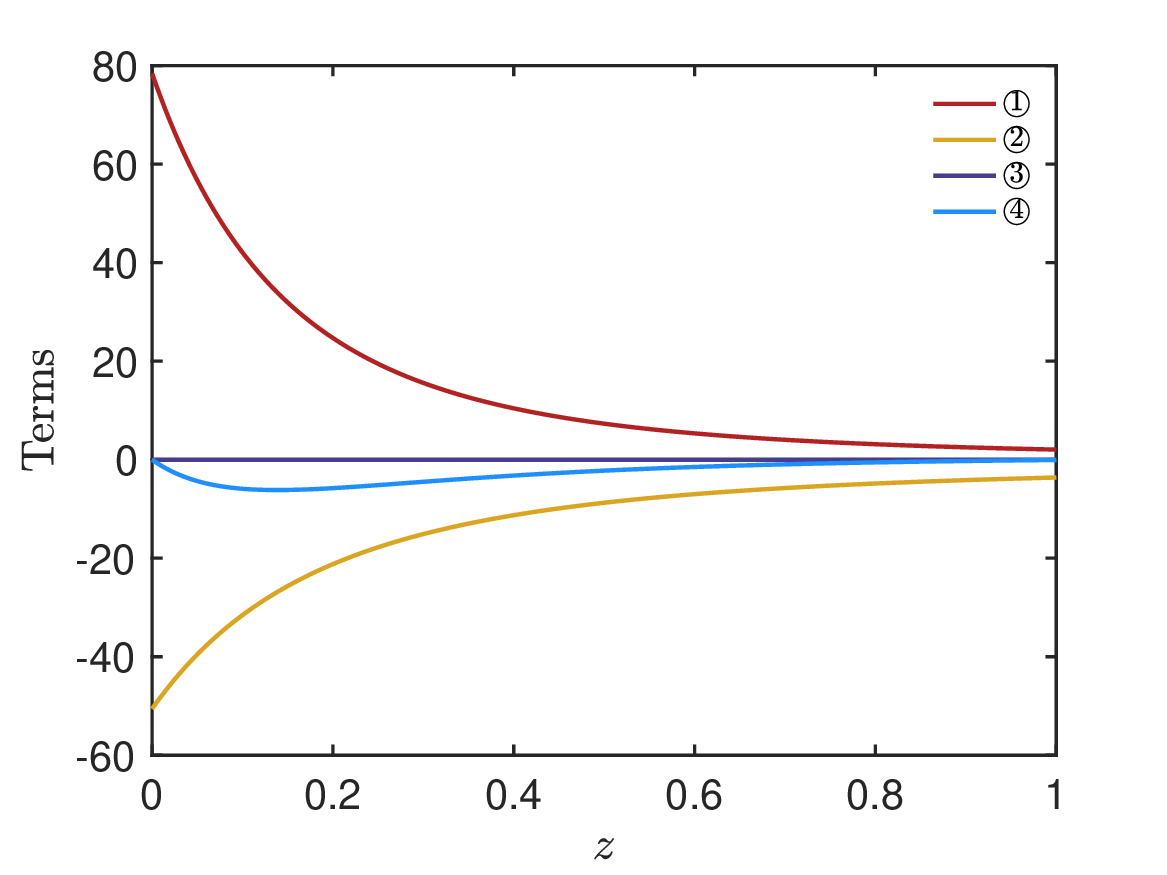}
				\put(3,72){(b)}
			\end{overpic}
		\end{minipage}
	}\\
	\subfigure {\
		\begin{minipage}[b]{.46\linewidth}
			\centering
			\begin{overpic}[scale=0.44]{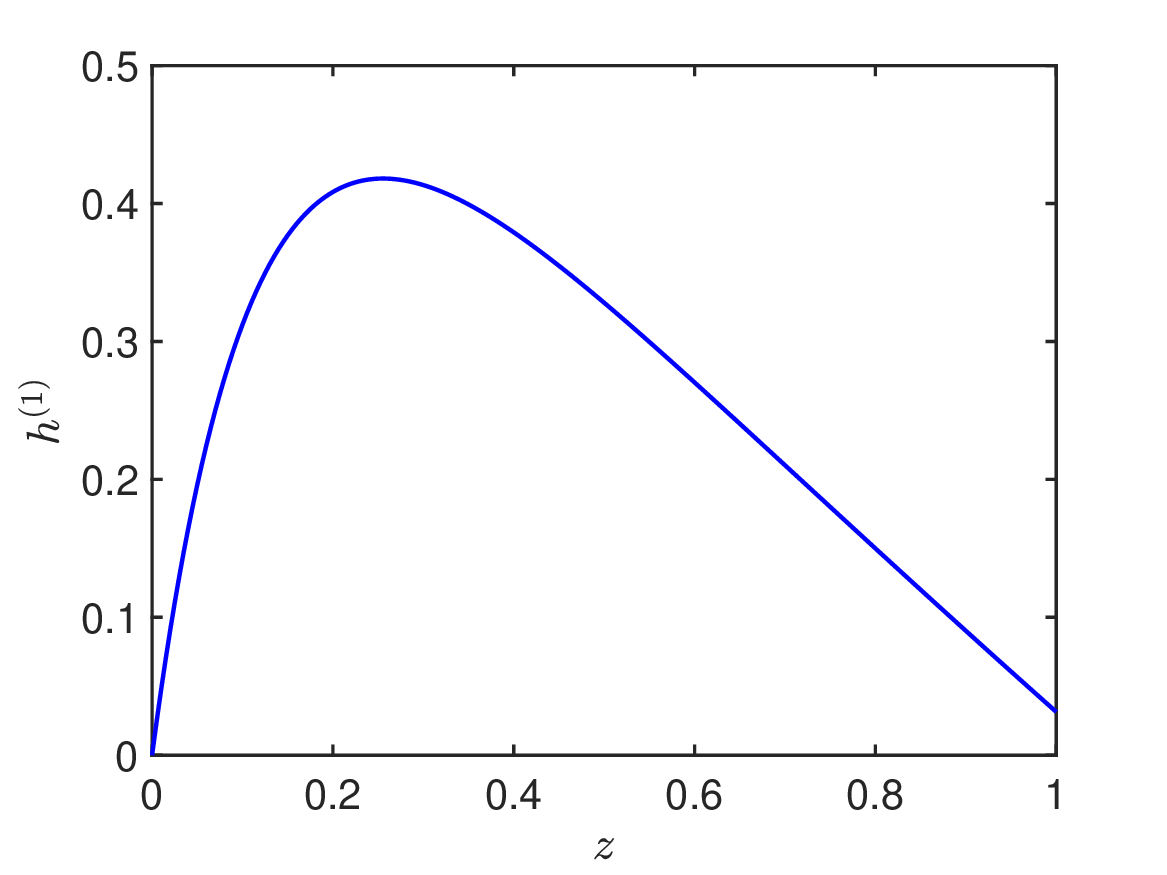}
				\put(3,72){(c)}
			\end{overpic}
		\end{minipage}
	}
	\caption{(a) The quantity $C^{(1)}$, (b) the terms $\textcircled{\small 1}-\textcircled{\small 4}$, and (c) $h^{(1)}$ versus $z$ for $\mathrm{D}=1.5,\ \mathrm{Ca}=1.8,\ \phi=0.95,\ \mathrm{Re}=0,\ \alpha=0,\  \beta=0.1$.}\label{fig:ss6}
\end{figure}

Term $\textcircled{\small 3}$ arises from the quadratic nonlinear terms in the Giesekus model. Term $\textcircled{\small 4}$ is similar to the second term in \eqref{3.22}, does not directly involve elastic stresses, and can be thought of as a response to the elastic mechanism. In order to determine which of the terms dominate we plot the sizes of each of the four terms in Figures~\ref{fig:ss5} and \ref{fig:ss6} for moderate hole size and large hole size, respectively.

In the case of moderate hole size, Figure~\ref{fig:ss5}(a) and (b) show that $\textcircled{\small 2}$ plays the dominant role in causing the integrand $C^{(1)}$ in \eqref{4.14} to be negative. Then the first-order second normal stress difference $T^{(1)}_2$ is negative and drives a radially inward flow. This, in turn, leads to $h^{(1)}(z=1)<0$, as shown in Figure~\ref{fig:ss5}(c). In summary, for moderate hole size, hole closure is enhanced by elastic effects due to radial advection of stresses arising from the surface-tension-driven flow.

In the case of large hole size, we observe opposite effects of elasticity on the hole size at the outlet, as illustrated in Figure~\ref{fig:phidiffDr} and Figure~\ref{fig:phidiffCa}, where $h^{(1)}(z=1)$ is positive. We also analyze the individual terms in \eqref{4.16} to explain this phenomenon, which is depicted in Figure~\ref{fig:ss6}. In contrast to Figure~\ref{fig:ss5}, we see that $h^{(1)}(z=1)>0$. Using \eqref{4.14}, we see that this must be the result of the integral of $C^{(1)}$ being positive. In Figure~\ref{fig:ss6}(a) we see that $C^{(1)}$ is large and positive for $z\lesssim0.15$ and is slightly negative for $z\gtrsim0.15$. Since the integral of $C^{(1)}$ is positive this means that the region $z\lesssim0.15$ is dominant. From Figure~\ref{fig:ss6}(b) we see that $\textcircled{\small 1}$ is the dominant term in the region $z\lesssim0.15$. We note that $\textcircled{\small 1}$ is extremely large near $z=0$ and becomes smaller as $z$ increases to $1$. This can be explained by $C^{(0)}_{,z}$, which appears in $\textcircled{\small 1}$ and using \eqref{3.9} is given by
\begin{align}
C^{(0)}_{,z}=\frac{h^{(0)2} H^{(0)}_{,z}-H^{(0)2} h^{(0)}_{,z}}{2 \mathrm{Ca} \big(H^{(0)}-h^{(0)}\big)^2}.\label{3.36}
\end{align}
Since the hole size is very large at the inlet, $H^{(0)}$ and $h^{(0)}$ are very close to each other near $z=0$. Using \eqref{3.36} this means that $C^{(0)}_{,z}$ is extremely large near $z=0$. As $z$ approaches 1, the difference between $H^{(0)}$ and $h^{(0)}$ increases, which results in $\textcircled{\small 1}$ becoming less positive. Therefore, $C^{(1)}$ is positive for $z\lesssim0.15$ and becomes negative for $z\gtrsim0.15$. We note that $\textcircled{\small 1}\propto u^{(0)}_0 C^{(0)}_{,z}>0$, and so the first-order second normal stress difference $T^{(1)}_2$ is positive. This drives a radially outward flow that suppresses hole closure. In summary, for large hole size, hole closure is suppressed by elastic effects due to axial advection of stresses arising from the surface-tension-driven flow. We note that these axial advection terms are only dominated near the inlet, and this effect can nevertheless be strong enough to dominate the overall hole closure.

\section{Linear stability analysis}\label{sec:5}
In this section, we carry out a linear stability analysis to determine the onset of draw resonance and obtain the critical draw ratio for different values of the parameters. To investigate the linear stability with a truncation error of $\mathcal{O}(\mathrm{De}^2)$, we denote the steady-state profiles in section~\ref{sec:3.3} by $\hat{\psi}=\hat{\psi}^{(0)}+\mathrm{De} \hat{\psi}^{(1)}$, where $\hat{\psi}^{(0)}$ and $\hat{\psi}^{(1)}$ represent the leading-order and first-order steady-state solutions for the quantities $\bm{\tau}_0,\ p_0,\ \mathbf{u}_0,\ h,\ H$ and $C$, which are given in section~\ref{sec:3.3}. We add small perturbations of the form $\tilde{\psi} e^{\omega t}=(\tilde{\psi}^{(0)}+\mathrm{De} \tilde{\psi}^{(1)})e^{\omega t}$, where $\tilde{\psi}^{(0)}$ and $\tilde{\psi}^{(1)}$ represent the leading-order and first-order perturbation amplitudes for the various quantities and $\omega$ is the growth rate of the perturbations. Hence, we take
\begin{align}
\psi=\hat{\psi}+\tilde{\psi}e^{\omega t}\quad \text{with}\quad \omega=\omega_0+\mathrm{De} \omega_1,\label{3.34}
\end{align}
where we have also expanded $\omega$ as a series in $\mathrm{De}$, $\omega_0$ being the growth rate for the Newtonian flow and $\omega_1$ representing the effect of elasticity on the growth rate.

Substituting \eqref{3.34} into \eqref{2.31}$-$\eqref{2.33}, linearizing for small perturbations, and equating terms of the same order in $\mathrm{De}$, one can represent $(\tilde{\tau}^{(0)}_{rr0},\ \tilde{\tau}^{(0)}_{\theta\theta 0},\ \tilde{\tau}^{(0)}_{zz0})$ and $(\tilde{\tau}^{(1)}_{rr0},\ \tilde{\tau}^{(1)}_{\theta\theta 0},\ \tilde{\tau}^{(1)}_{zz0})$ in terms of $\tilde{u}^{(0)}_0$ and $\tilde{u}^{(1)}_0$. Similarly, one can express $(\tilde{p}^{(0)}_0,\ \tilde{C}^{(0)})$ and $(\tilde{p}^{(1)}_0,\ \tilde{C}^{(1)})$ in terms of $(\tilde{u}^{(0)}_0,\ \tilde{h}^{(0)},\ \tilde{H}^{(0)})$ and $(\tilde{u}^{(1)}_0,\ \tilde{h}^{(1)},\ \tilde{H}^{(1)})$. By substituting the perturbed quantities \eqref{3.34} into \eqref{2.38}$-$\eqref{2.40} and linearizing for small perturbations, we obtain a general eigenvalue problem for the perturbations $\tilde{u}_0,\ \tilde{h}$, and $\tilde{H}$. The detailed expressions of these equations are omitted here for the sake of brevity. Actually, this eigenvalue problem can easily be derived using a symbolic manipulation platform such as \textit{Mathematica}. The boundary conditions for the general perturbed eigenvalue problem are 
\begin{align}
&\tilde{u}^{(0)}=0,\quad \tilde{u}^{(1)}=0\quad \text{at}\quad z=0\quad \text{and}\quad z=1.\label{3.34a}\\
&\tilde{h}^{(0)}=0,\quad \tilde{h}^{(1)}=0,\quad \tilde{H}^{(0)}=0,\quad \tilde{H}^{(1)}=0\quad \text{at}\quad z=0.\label{3.35a}
\end{align}
At leading order in $\mathrm{De}$, the eigenvalue problem can be written in the form $(\tilde{\bm{L}}-\omega_0) \tilde{\bm{x}}^{(0)}=0$, where $\tilde{\bm{x}}^{(0)}=(\tilde{u}^{(0)}_0,\ \tilde{h}^{(0)},\ \tilde{H}^{(0)})^\intercal$, and $\tilde{\bm{L}}$ is a linear differential operator whose coefficients depend on the steady-state profiles. This eigenvalue problem recovers the stability results for a Newtonian fluid. Furthermore, at first order in $\mathrm{De}$, we obtain an equation of the form $(\tilde{\bm{L}}-\omega_0) \tilde{\bm{x}}^{(1)}=\bm{G}(\omega_1, \tilde{\bm{x}}^{(0)})$, where $\bm{G}(\omega_1, \tilde{\bm{x}}^{(0)})$ is a function of $\omega_1$ and $\tilde{\bm{x}}^{(0)}$, and $\tilde{\bm{x}}^{(1)}=(\tilde{u}^{(1)}_0,\ \tilde{h}^{(1)},\ \tilde{H}^{(1)})^\intercal$. The determination of $\omega_1$ involves utilizing the Fredholm alternative.

Defining the inner product as the integral over the domain and using the boundary conditions \eqref{3.34a}$-$\eqref{3.35a}, we introduce the adjoint problem of the leading-order eigenvalue problem, which we denote as
\begin{align}
(\tilde{\bm{L}}^+-\omega_0) \tilde{\bm{x}}^{+}=0,\label{A.3}
\end{align}
where $\tilde{\bm{L}}^+$ is the adjoint operator of $\tilde{\bm{L}}$. By taking the inner product of the first-order equation with $\tilde{\bm{x}}^{+}$, we obtain
\begin{align}
\langle  \tilde{\bm{x}}^{+}, \;  (\tilde{\bm{L}}-\omega_0) \tilde{\bm{x}}^{(1)}\rangle =\langle  \tilde{\bm{x}}^{+}, \; \bm{G}(\omega_1, \tilde{\bm{x}}^{(0)})  \rangle.\label{A.4}
\end{align}
We then use integration by parts along with \eqref{A.3} to obtain
\begin{align}
\langle  \tilde{\bm{x}}^{+}, \; \bm{G}(\omega_1, \tilde{\bm{x}}^{(0)})  \rangle=0, \label{A.5}
\end{align}
which represents the solvability condition for $\omega_1$. 

The leading-order eigenvalue problem can be discretized using a spectral collocation method, resulting in a matrix eigenvalue problem. By employing a sufficiently large number of grid points, we can obtain a reliable estimate of the eigenvalue spectrum for the operator. Since we are mostly concerned to know if the solution is stable or not, our focus lies on identifying the eigenvalue with the largest real part and its corresponding eigenfunction. 

Having obtained the numerical solution to the leading-order problem, \eqref{A.5} can also be readily solved using a similar spectral collocation method. The quantity $\omega_1$ represents the difference in growth rates between the viscoelastic and Newtonian fluids. However, our main concern is to determine if the real part of $\omega=\omega_0+\mathrm{De} \omega_1$ is positive or negative. Given our assumption that $\mathrm{De}\ll1$, $\omega_1$ can only make a significant difference in determining if the solution is stable or not for the values of the parameters for which the real part of $\omega_0$ is close to zero. Hence, if the real part of $\omega_1$ is positive, then elastic effects act to make the flow more unstable or less stable than the Newtonian flow. We denote the critical draw ratio derived from the leading-order eigenvalue problem as $\mathrm{D}_{0c}$, which signifies the onset of instability in the case of a Newtonian fluid. By analyzing the magnitude of $\omega_1$ at $\mathrm{D}_{0c}$, we can determine whether elasticity stabilizes or destabilizes the drawing process.
\begin{figure}
	\centering
	\subfigure {\
		\begin{minipage}[b]{.46\linewidth}
			\centering
			\begin{overpic}[scale=0.44]{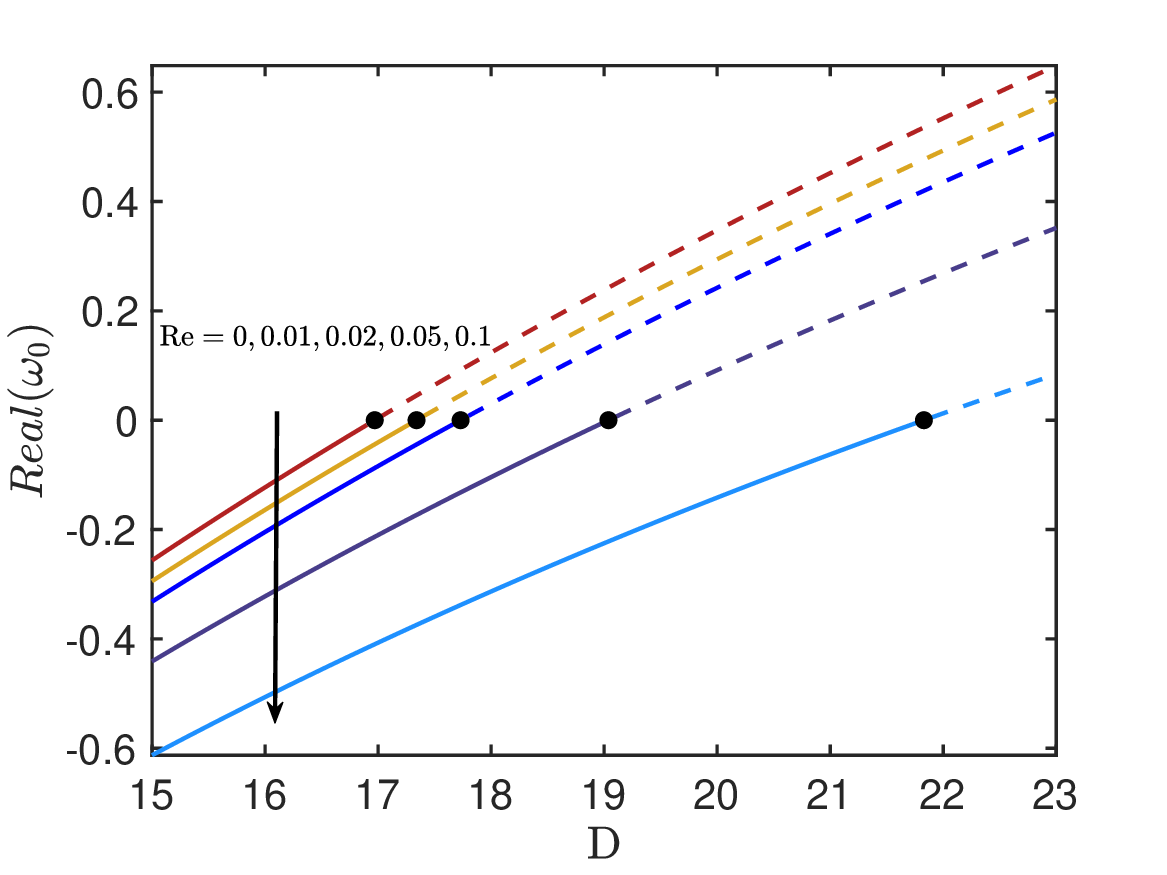}
				\put(3,72){(a)}
			\end{overpic}
		\end{minipage}
	}
	\subfigure {\
		\begin{minipage}[b]{.46\linewidth}
			\centering
			\begin{overpic}[scale=0.44]{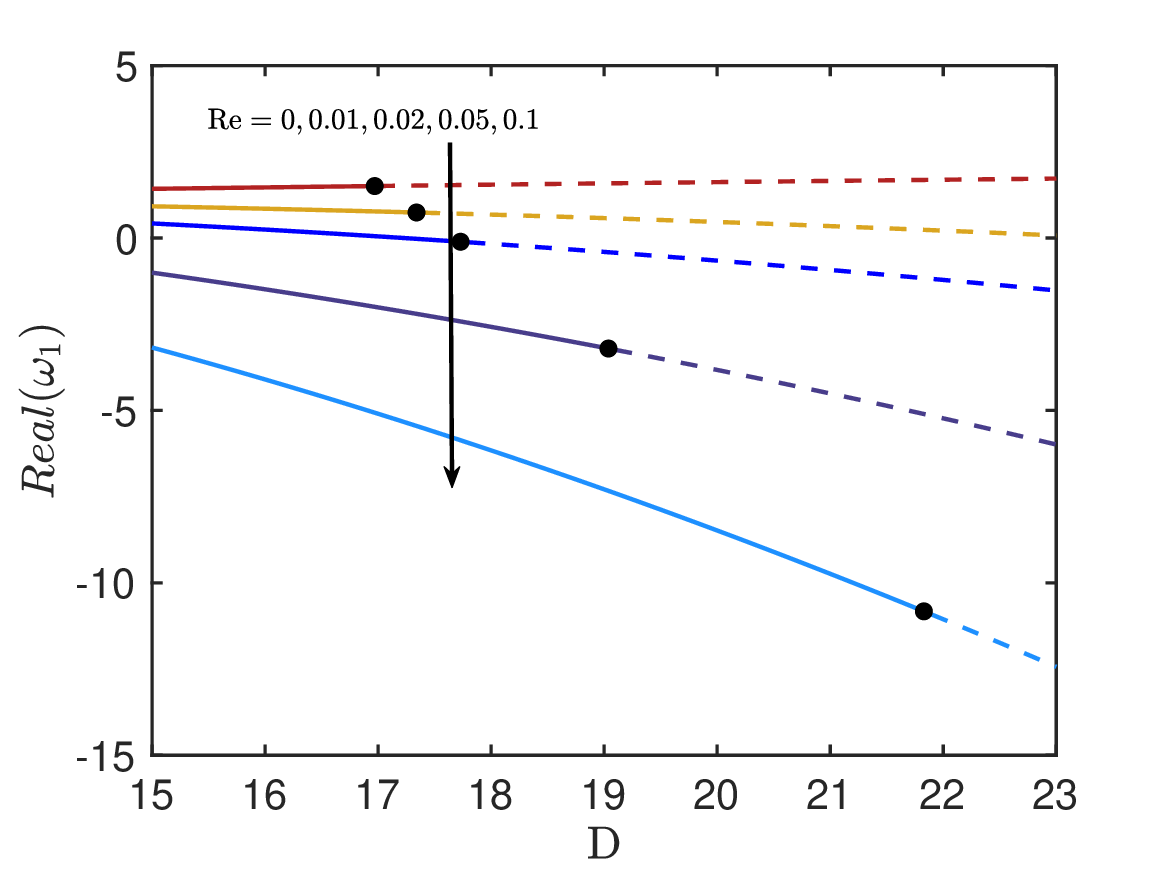}
				\put(3,72){(b)}
			\end{overpic}
		\end{minipage}
	}\caption{(a) $Real(\omega_0)$ versus draw ratio $\mathrm{D}$ for various values of the Reynolds number $\mathrm{Re}$. (b) The elastic contribution to the growth rate $Real(\omega_1)$ versus $\mathrm{D}$ for various $\mathrm{Re}$. Other parameters are $\mathrm{Ca}=2,\ \phi=0.5,\ \alpha=0,\ \beta=0$. For each curve, there is a value of $\mathrm{D}$ marked by a black circle at which $Real(\omega_0)=0$, below which (solid curve) the Newtonian flow is stable ($Real(\omega_0)<0$), above which (dashed curve) it is unstable ($Real(\omega_0)>0$). If $Real(\omega_0+\mathrm{De}\  \omega_1)=Real(\omega_0)+\mathrm{De}\ Real(\omega_1)>0$, the draw process is unstable.}
	\label{fig:sa1}
\end{figure}

In Figure~\ref{fig:sa1}(a), curves of the growth rates $Real(\omega_0)$ for Newtonian fluids are shown for different values of Reynolds number $\mathrm{Re}$. For each Reynolds number, the critical draw ratio $\mathrm{D}_{0c}$ is at the point of intersection of the curve with the line $Real(\omega_0)=0$ (black circle), the curve is solid for $Real(\omega_0)<0$ (stable) and dashed for $Real(\omega_0)>0$ (unstable). Similarly, in Figure~\ref{fig:sa1}(b) we plot $Real(\omega_1)$, which measures how elasticity affects the growth rate, versus the draw ratio $\mathrm{D}$. Since $\mathrm{De}\ll1$, the most important value of $Real(\omega_1)$ corresponds to the draw ratio $\mathrm{D}_{0c}$ at which $Real(\omega_0)=0$. This is marked as a solid black circle in Figure~\ref{fig:sa1}(b). In Figure~\ref{fig:sa1}(b) the solid curve corresponds to values of $\mathrm{D}$ for which the zero $\mathrm{De}$ flow is stable ($Real(\omega_0)<0$) and the dashed curve corresponds to values of $\mathrm{D}$ for which the zero $\mathrm{De}$ flow is unstable ($Real(\omega_0)>0$). At the critical draw ratio $\mathrm{D}_{0c}$ for the zero $\mathrm{De}$ flow (corresponding to the black circles in Figures~\ref{fig:sa1}(a) and (b)), we see that $Real(\omega_1)$ is positive in the cases of $\mathrm{Re}=0$ and $\mathrm{Re}=0.01$, indicating that elastic effects destabilize the drawing process. Conversely, when $\mathrm{Re}=0.02,\ 0.05,$ and $\mathrm{Re}=0.1$, elastic effects stabilize the process, since $Real(\omega_1)<0$.

\begin{figure}
	\centering
	\subfigure {\
		\begin{minipage}[b]{.46\linewidth}
			\centering
			\begin{overpic}[scale=0.44]{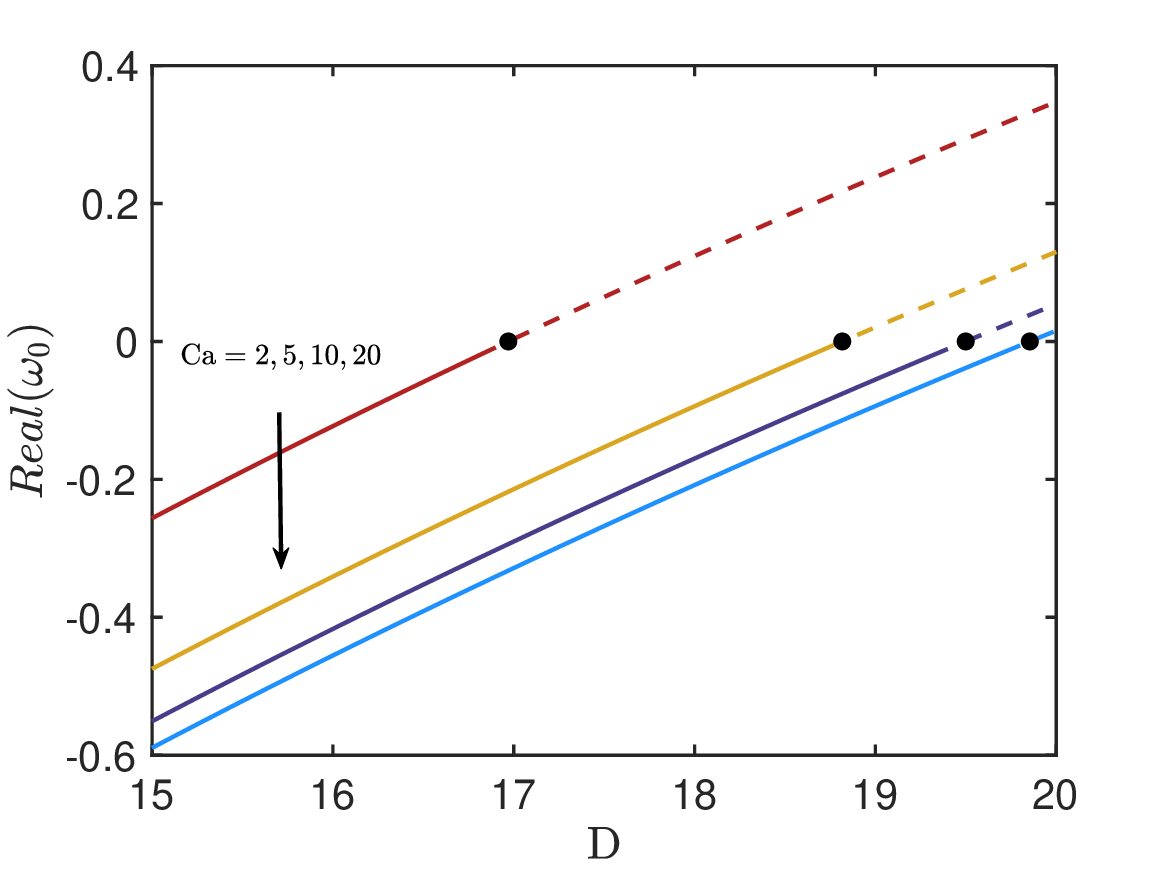}
				\put(3,72){(a)}
			\end{overpic}
		\end{minipage}
	}
	\subfigure {\
		\begin{minipage}[b]{.46\linewidth}
			\centering
			\begin{overpic}[scale=0.44]{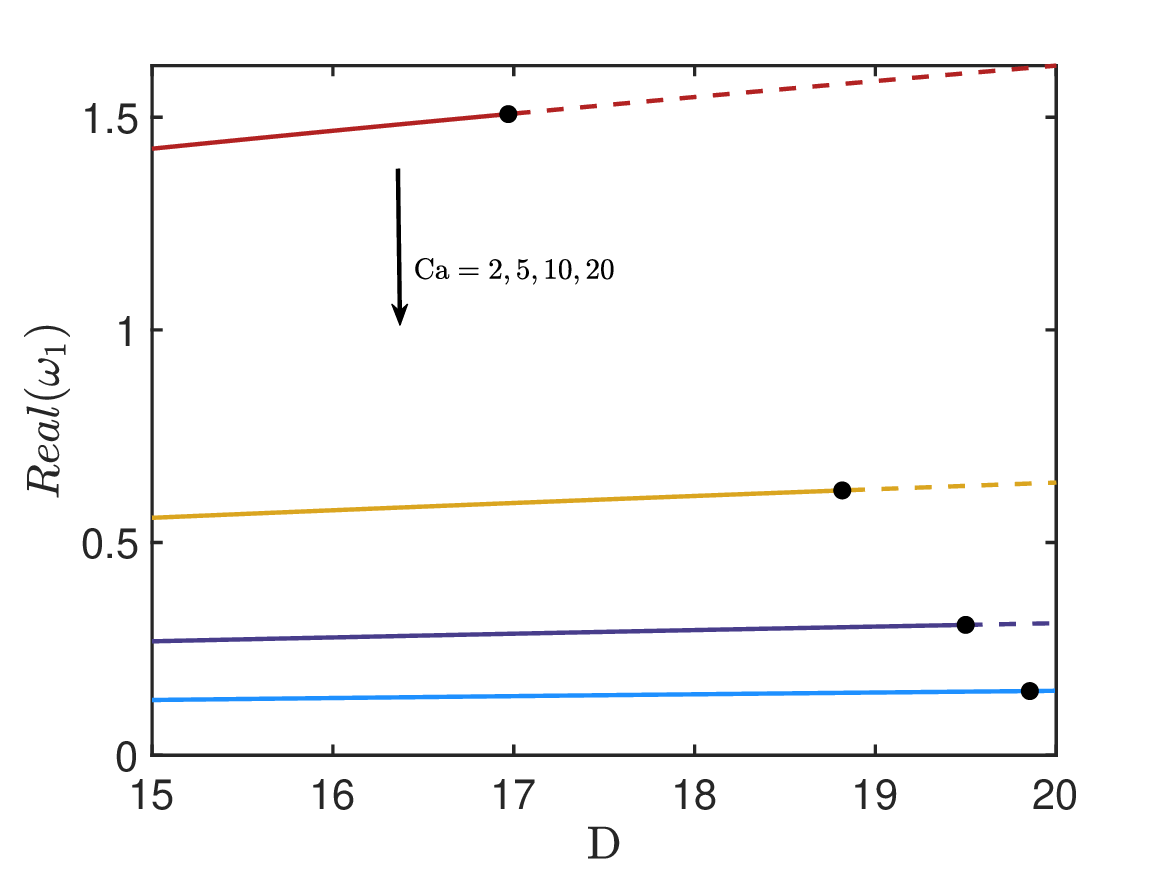}
				\put(3,72){(b)}
			\end{overpic}
		\end{minipage}
	}\\
	\subfigure {\
		\begin{minipage}[b]{.46\linewidth}
			\centering
			\begin{overpic}[scale=0.44]{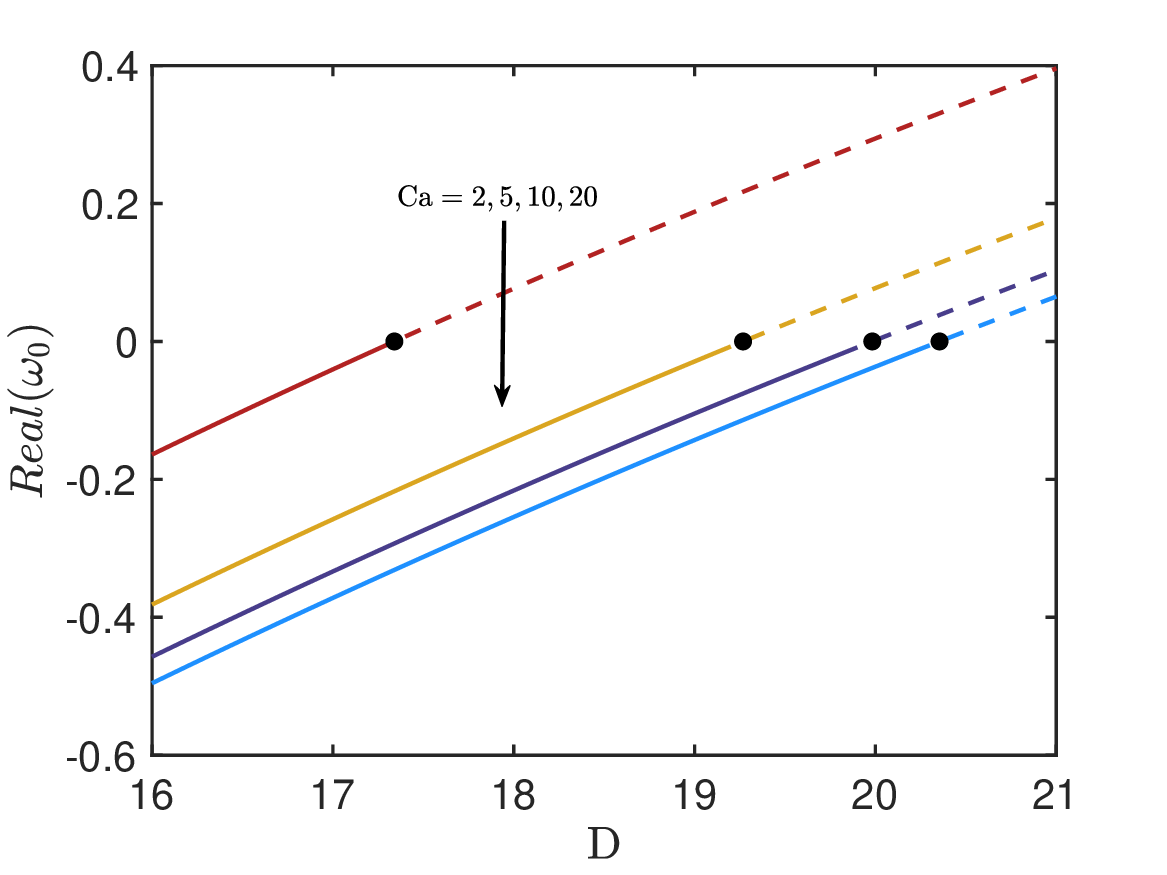}
				\put(3,72){(c)}
			\end{overpic}
		\end{minipage}
	}
	\subfigure {\
		\begin{minipage}[b]{.46\linewidth}
			\centering
			\begin{overpic}[scale=0.44]{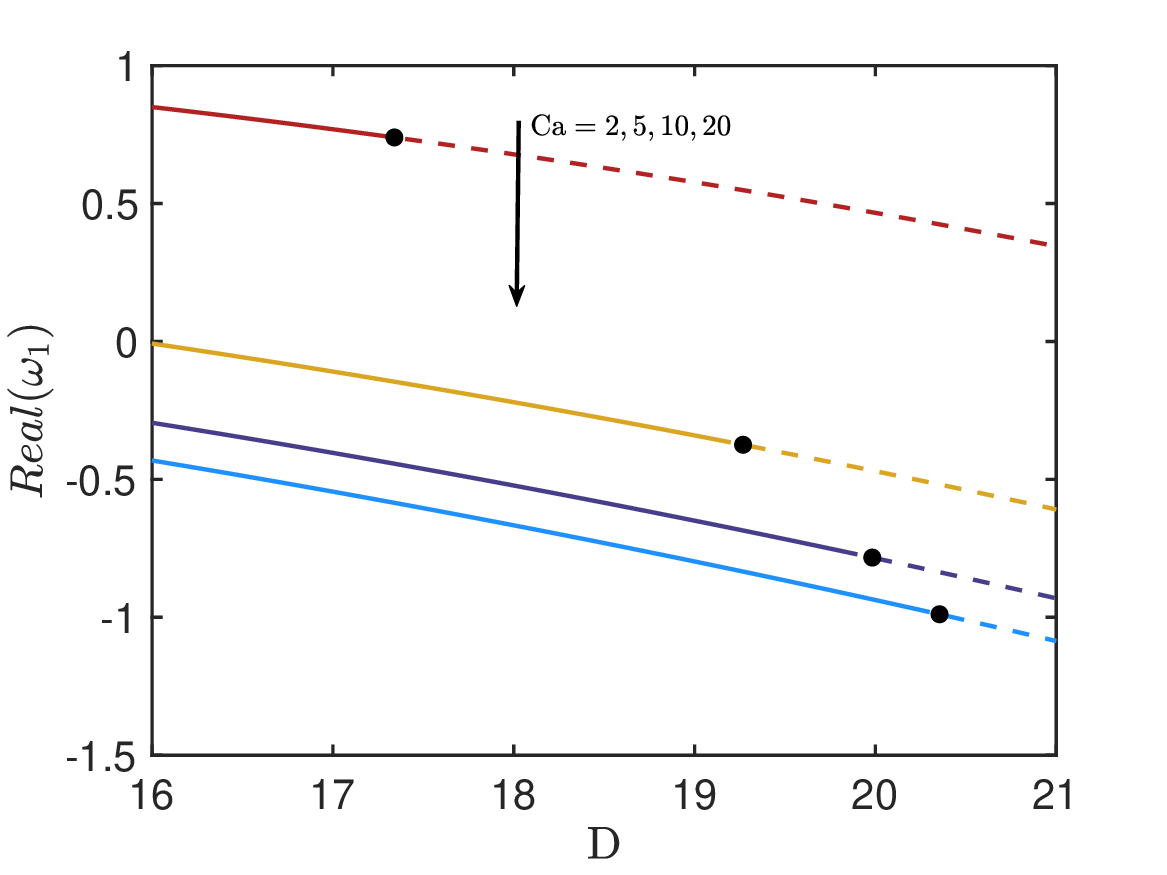}
				\put(3,72){(d)}
			\end{overpic}
		\end{minipage}
	}
	\caption{$Real(\omega_0)$ versus draw ratio $\mathrm{D}$ for various values of the capillary number $\mathrm{Ca}$ for (a) $\mathrm{Re}=0$ and (c) $\mathrm{Re}=0.01$, respectively. The elastic contribution to growth rate $Real(\omega_1)$ versus $\mathrm{D}$ for various $\mathrm{Ca}$ for (b) $\mathrm{Re}=0$ and (d) $\mathrm{Re}=0.01$, respectively. Other parameters are $\phi=0.5,\ \alpha=0,\ \beta=0$. For each curve, there is a value of $\mathrm{D}$ marked by a black circle at which $Real(\omega_0)=0$, below which (solid curve) the Newtonian flow is stable ($Real(\omega_0)<0$), above which (dashed curve) it is unstable ($Real(\omega_0)>0$). If $Real(\omega_0+\mathrm{De}\  \omega_1)=Real(\omega_0)+\mathrm{De}\ Real(\omega_1)>0$, the draw process is unstable.}\label{fig:sa3}
\end{figure}

Similar plots are shown in Figure~\ref{fig:sa3} for fixed $\mathrm{Re}$ and different values of the capillary number $\mathrm{Ca}$. Figure~\ref{fig:sa3}(a) and (b) correspond to $\mathrm{Re}=0$, while Figure~\ref{fig:sa3}(c) and (d) correspond to $\mathrm{Re}=0.01$. Figure~\ref{fig:sa3}(b) reveals that elastic effects consistently destabilize the process at $\mathrm{Re}=0$ for all $\mathrm{Ca}$ considered. On the other hand, Figure~\ref{fig:sa3}(d) shows that elastic effects destabilize the process if $\mathrm{Ca}=2$, which is consistent with the findings in Figure~\ref{fig:sa1}. However, for $\mathrm{Ca}\geq 5$ elastic effects stabilize the drawing process.

\begin{figure}
	\centering
	\subfigure {\
		\begin{minipage}[b]{.46\linewidth}
			\centering
			\begin{overpic}[scale=0.44]{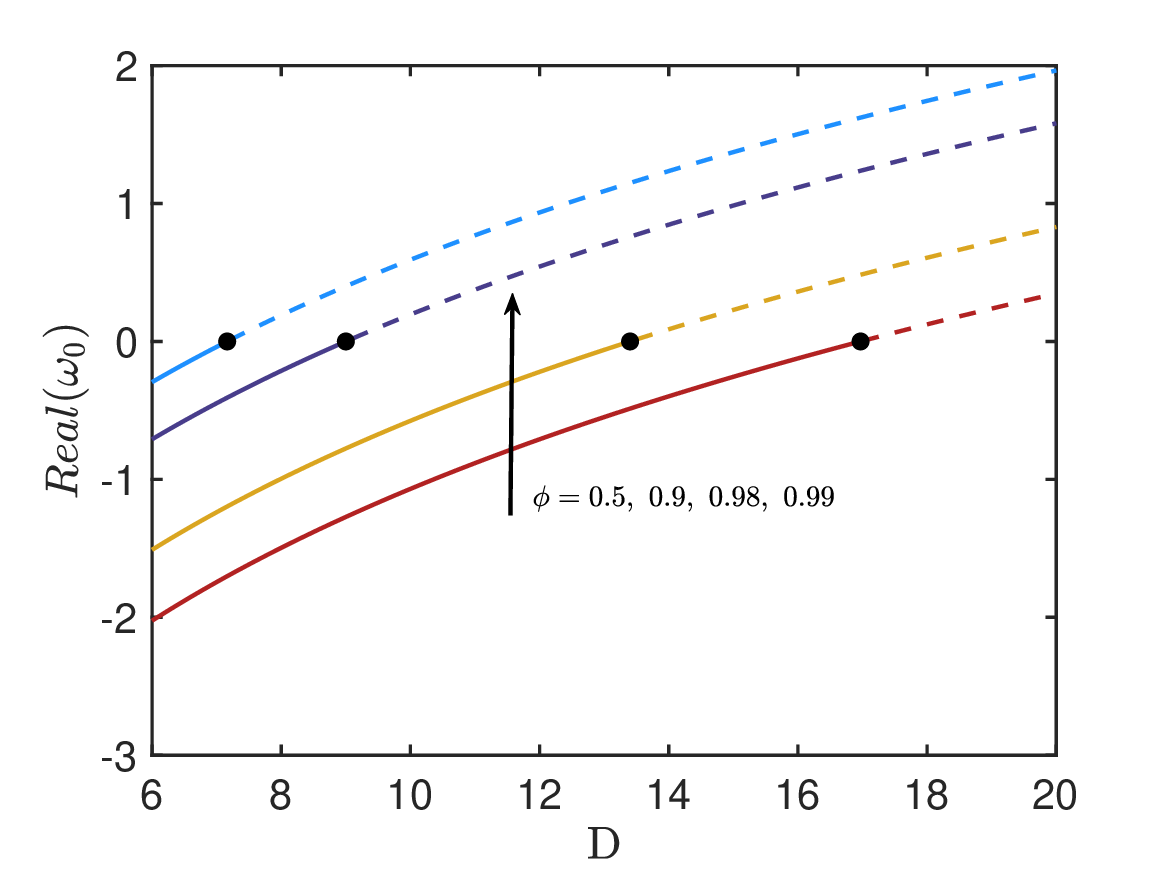}
				\put(3,72){(a)}
			\end{overpic}
		\end{minipage}
	}
	\subfigure {\
		\begin{minipage}[b]{.46\linewidth}
			\centering
			\begin{overpic}[scale=0.44]{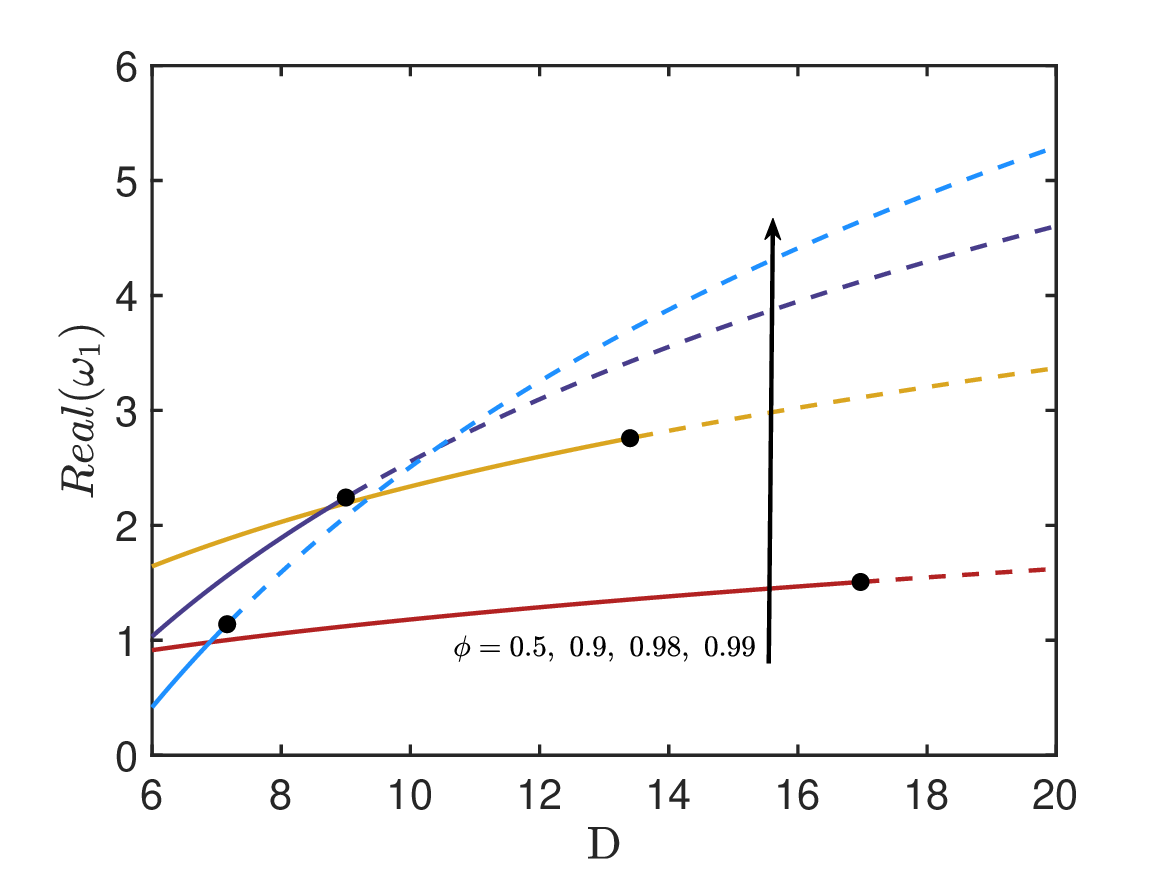}
				\put(3,72){(b)}
			\end{overpic}
		\end{minipage}
	}\\
	\subfigure {\
		\begin{minipage}[b]{.46\linewidth}
			\centering
			\begin{overpic}[scale=0.44]{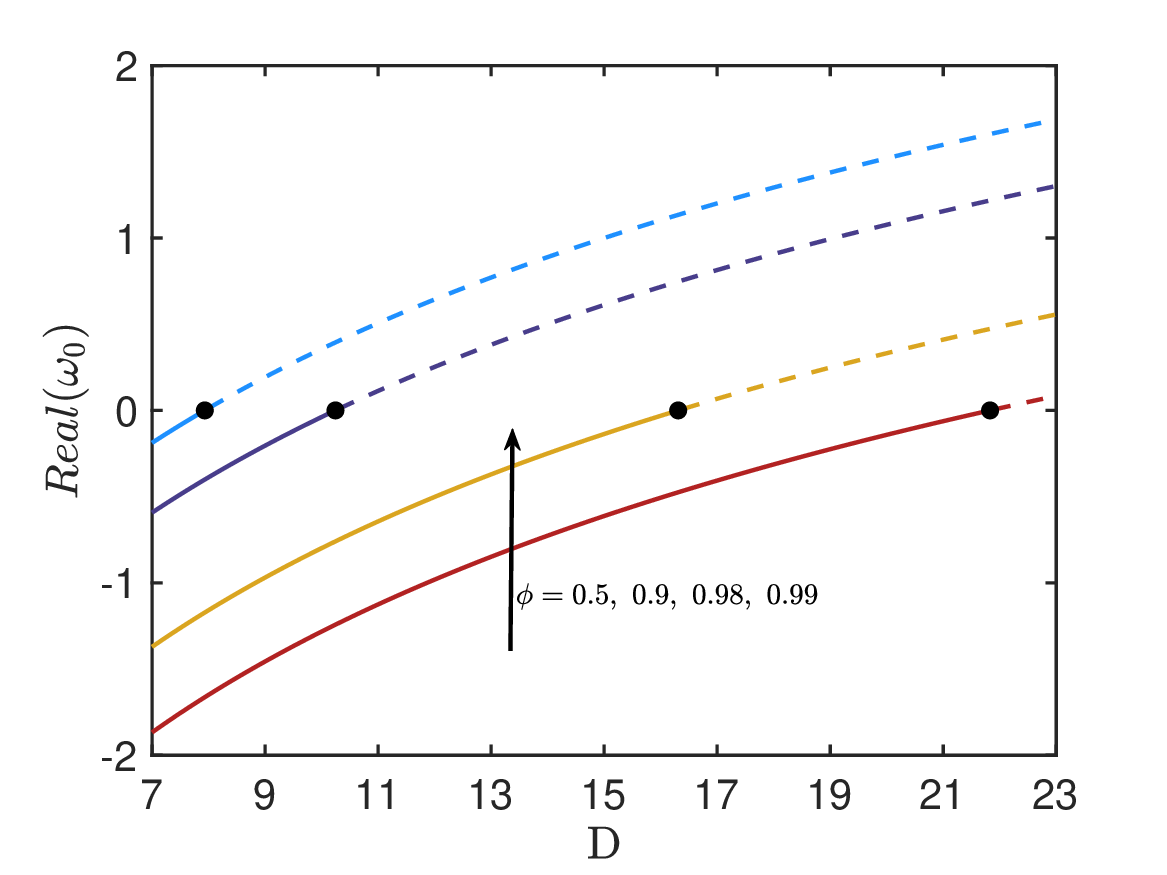}
				\put(3,72){(c)}
			\end{overpic}
		\end{minipage}
	}
	\subfigure {\
		\begin{minipage}[b]{.46\linewidth}
			\centering
			\begin{overpic}[scale=0.44]{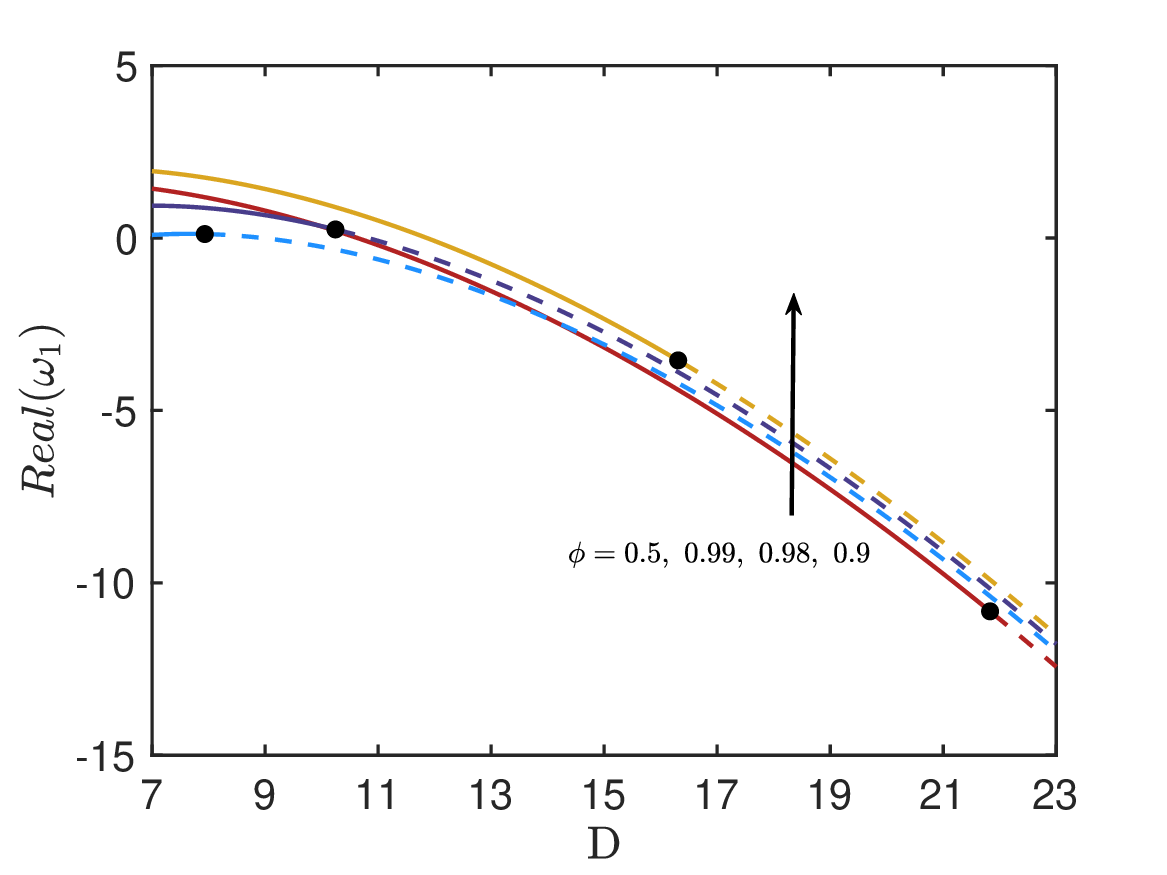}
				\put(3,72){(d)}
			\end{overpic}
		\end{minipage}
	}
	\caption{$Real(\omega_0)$ versus draw ratio $\mathrm{D}$ for various values of $\phi$ for (a) $\mathrm{Re}=0$ and (c) $\mathrm{Re}=0.1$, respectively. The elastic contribution to growth rate $Real(\omega_1)$ versus $\mathrm{D}$ for various $\phi$ with (b) $\mathrm{Re}=0$ and (d) $\mathrm{Re}=0.1$, respectively. Other parameters are $\mathrm{Ca}=2,\ \alpha=0,\ \beta=0$. For each curve, there is a value of $\mathrm{D}$ marked by a black circle at which $Real(\omega_0)=0$, below which (solid curve) the Newtonian flow is stable ($Real(\omega_0)<0$), above which (dashed curve) it is unstable ($Real(\omega_0)>0$). If $Real(\omega_0+\mathrm{De}\  \omega_1)=Real(\omega_0)+\mathrm{De}\ Real(\omega_1)>0$, the draw process is unstable.}\label{fig:sa5}
\end{figure}
In Figure~\ref{fig:sa5}(a) and (c), the growth rates $Real(\omega_0)$ for Newtonian fluids are shown with different inlet hole sizes, for $\mathrm{Re}=0$ and $\mathrm{Re}=0.1$, respectively. Again, the critical draw ratios for Newtonian fluids, corresponding to different $\phi$, are observed at the intersections of the curves with the line $Real(\omega_0)=0$ (black circles), with the solid and dashed portions of the curves denoting stability/instability of the zero $\mathrm{De}$ flow. Furthermore, the influence of inlet hole size on $Real(\omega_1)$ is shown in Figures~\ref{fig:sa5}(b) and (d), for varying values of $\phi$. We find that $Real(\omega_1)$ is always positive if $\mathrm{Re}=0$, indicating that elasticity destabilizes the drawing process. However, when $\mathrm{Re}=0.1$, elastic effects stabilize the process except for large values of $\phi$ very close to unity, in which case it becomes destabilizing.

\begin{figure}
	\centering
	\subfigure {\
		\begin{minipage}[b]{.46\linewidth}
			\centering
			\begin{overpic}[scale=0.44]{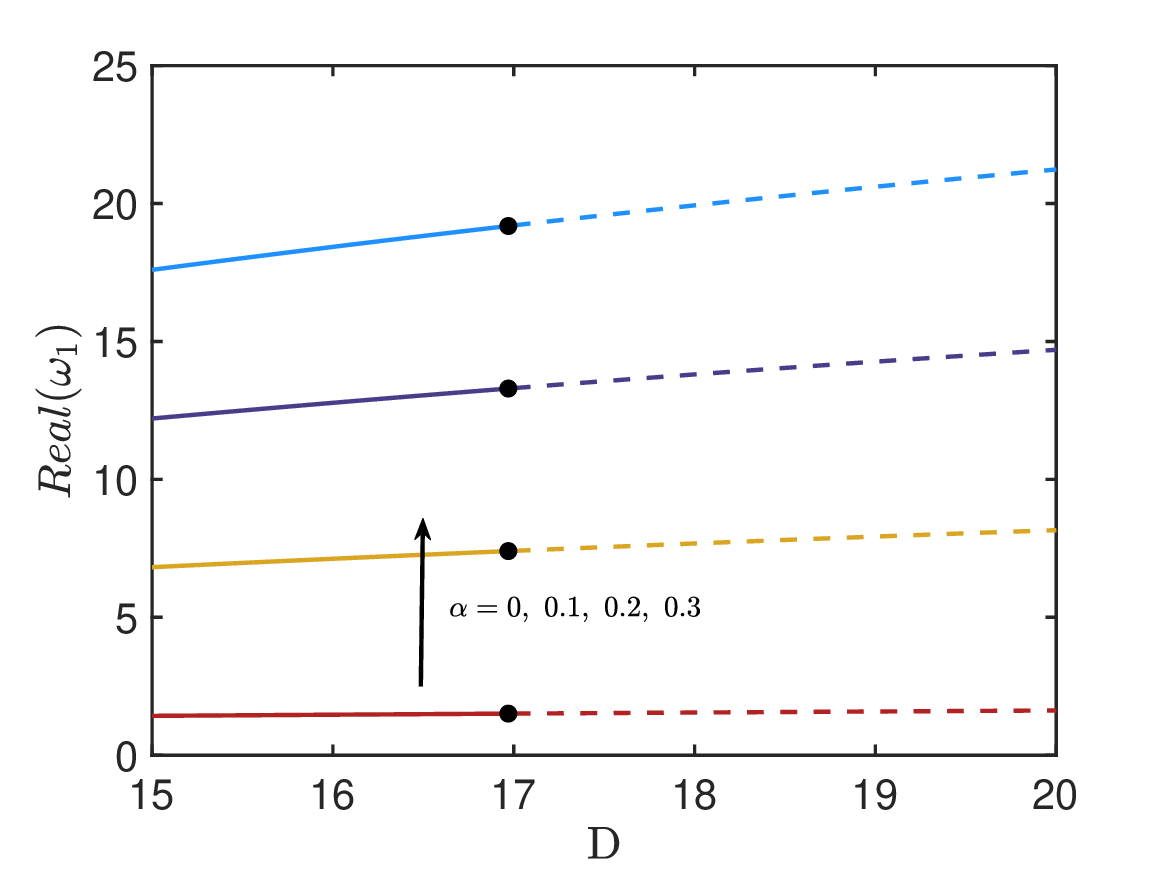}
				\put(3,72){(a)}
			\end{overpic}
		\end{minipage}
	}
	\subfigure {\
		\begin{minipage}[b]{.46\linewidth}
			\centering
			\begin{overpic}[scale=0.44]{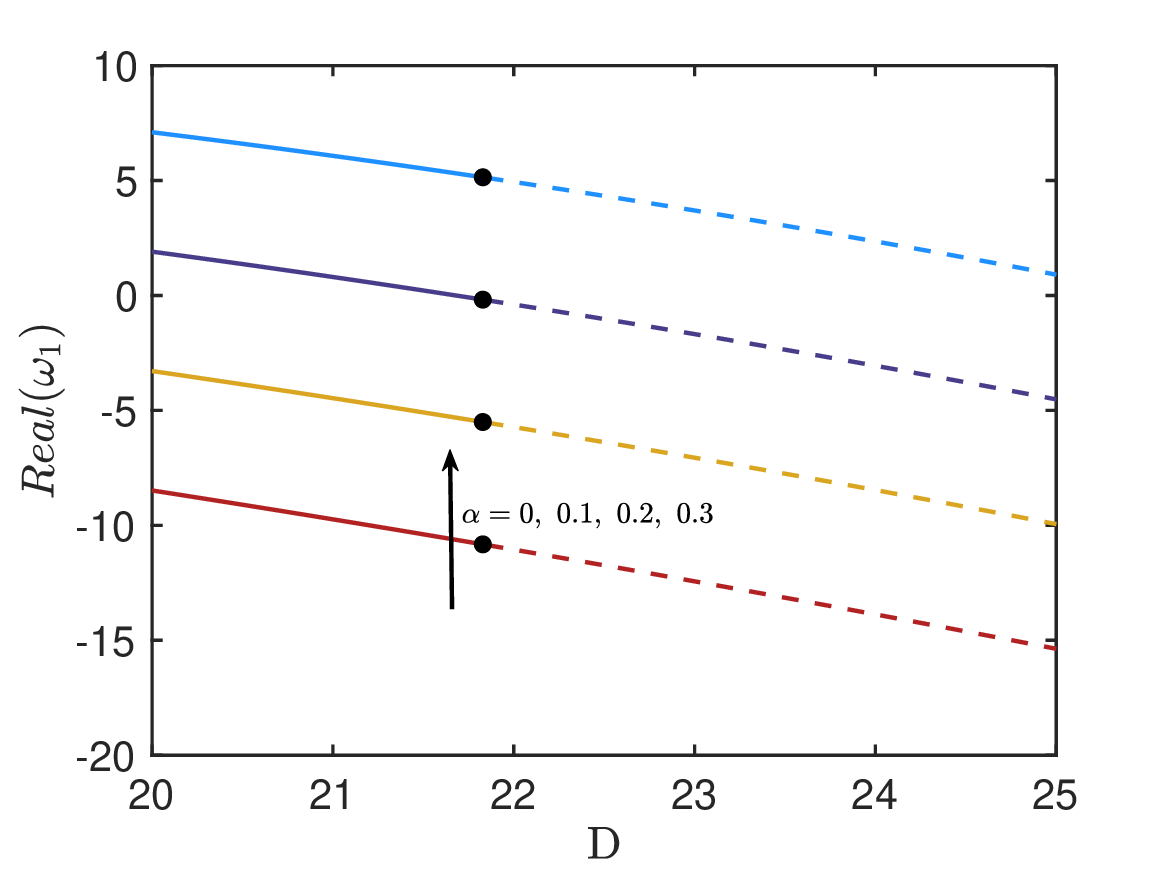}
				\put(3,72){(b)}
			\end{overpic}
		\end{minipage}
	}
	\caption{$Real(\omega_1)$ versus draw ratio $\mathrm{D}$ for various values of the mobility factor $\alpha$ for (a) $\mathrm{Re}=0$ and (b) $\mathrm{Re}=0.1$. Other parameters are $\phi=0.5,\ \mathrm{Ca}=2,\ \beta=0.$ Besides, with these parameters, $Real(\omega_0)$ are plotted against $\mathrm{D}$ in Figure~\ref{fig:sa1}(a).}\label{fig:sa4}
\end{figure}

In Figure~\ref{fig:sa4}(a) and (b), we examine the impact of the mobility factor $\alpha$ on the stability of the drawing process for viscoelastic fluids at $\mathrm{Re}=0$ and $\mathrm{Re}=0.1$, respectively. We note that at leading order in $\mathrm{De}$ the results for linear stability do not depend on $\alpha$ since this case represents a Newtonian fluid. Therefore, we do not plot $Real(\omega_0)$ versus $\mathrm{D}$, but just note from Figure~\ref{fig:sa1}(a) that $\mathrm{D}_{0c}\approx17$ for $\mathrm{Re}=0$ and $\mathrm{D}_{0c}\approx21.8$ for $\mathrm{Re}=0.1$. With $\alpha=0$, Figure~\ref{fig:sa4} reveals that elastic effects consistently destabilize the process if $\mathrm{Re}=0$, while they stabilize the process if $\mathrm{Re}=0.1$, which is consistent with the findings in previous figures. Additionally, we observe that the process becomes increasingly less stable as $\alpha$ increases, for both $\mathrm{Re}=0$ and $\mathrm{Re}=0.1$. When $\mathrm{Re}=0.1$, the elasticity plays a role to stabilize the process for $\alpha\leq0.2$, however, it destabilizes the process for larger $\alpha$.

The effect of the polymer concentration, $\beta$, on $Real(\omega_1)$ is not presented. This is because the role it plays is rather straightforward. As $\beta$ increases to $1$, the polymer concentration becomes zero, causing the flow to behave like a Newtonian fluid. Consequently, $Real(\omega_1)$ becomes zero when $\beta=1$. In fact as $\beta$ increases, $Real(\omega_1)$ monotonically tends to zero and the effect of elasticity is, hence, reduced.

\section{Conclusions}\label{sec:6}
In this paper, we have determined the steady-state behaviour and analyzed the stability of the drawing process for an axisymmetric tube composed of a Giesekus fluid. Our analysis takes into account surface tension and inertial effects. We have used asymptotic techniques to obtain a set of one-dimensional equations that describe the flow in the case of weak elastic effects. We have obtained the steady-state profiles and found that, in almost all of the parameter space, elastic effects make the hole close more rapidly than in the zero Deborah number case, equivalent to Newtonian fibre drawing. This is counterintuitive in the sense that one would probably naively imagine that the elastic stresses would oppose the inward radial flow induced by the surface tension and hence cause the hole to close more slowly. From our analysis, the stresses are advected by the zero Deborah number flow in both axial and radial directions. The advection causes changes in the stresses that will be opposed by elastic effects and hence induces corrections to the second normal stress difference. These second normal stress differences are the key mechanism by which the elasticity affects hole evolution. For almost the whole parameter space, the correction to the second normal stress difference is negative, which induces an additional radially inward flow that causes the hole to close more rapidly than in the Newtonian case. We also showed that there is a small region of the parameter space for which the hole radius at the inlet is very close to the outer radius. In this case, the axial advection of stress by the Newtonian flow causes the correction to the second normal stress difference to be large and positive near the inlet of the device, which induces a strong outward flow.  This outward flow near the inlet dominates the hole closure that occurs over the rest of the device, so that the hole at the device outlet is larger than in the Newtonian case.

In fibre drawing, a reduction of the hole closure due to surface tension effects is, typically, desired. Our results indicate that elastic effects will be detrimental to achieving this goal for tubes that have a moderate hole size at the inlet of the device. However, for tubes that have a very large hole size at the inlet of the device and sufficiently small draw ratio we do indeed obtain the desired effect. Moreover, there is a class of microstructured optical fibres (MOFs) that have very large air holes and a very small amount of their cross-sectional area occupied by the fluid \citep{argyros2007hollow}. These types of MOFs are very heavily affected by surface tension and it therefore seems that elastic effects may be very useful in the fabrication of such fibers.

We also examined the linear stability of the drawing process for weakly viscoelastic materials to determine whether elastic effects act to stabilize or destabilize the drawing process relative to the Newtonian case. We have shown that if inertia is negligible, elastic effects are always destabilizing. This is in direct contrast to the case of a solid thread described by the upper-convected Maxwell model for which elastic effects are stabilizing in the absence of inertia and surface tension \citep{zhou2010thermal}. This shows that surface tension and the presence of the hole fundamentally changes the stability characteristics of drawing. On the other hand, for non-zero inertia, elastic effects can be stabilizing if the capillary number is sufficiently large, or if the initial hole size is sufficiently small. The main goal in manufacturing is to suppress hole closure and our steady-state results showed that this only occurs for large input hole sizes. Unfortunately, for large input hole sizes elastic effects act to destabilize the flow. However, there is a trade-off between Reynolds number and hole size in determining stability, and inertia acts to stabilize. Elastic effects also act to stabilize the flow for small values of $\alpha$ indicating that Oldroyd-B fluids are more stable than Giesekus fluids. This will be of interest when selecting materials for drawing.

When drawing optical fibres, the draw ratio $\mathrm{D}$ is typically chosen to be as large as possible. In this case elastic effects enhance hole closure, which is not typically desirable in manufacturing settings. But for extrusion flows which typically have moderate $\mathrm{D}$, elastic effects can suppress hole closure for tubes with large air holes. Extrusion flows typically have very low $\mathrm{Re}$, so elastic effects will act to destabilize. But in extrusion flows draw ratios typically used in manufacturing settings are sufficiently small that the zero $\mathrm{De}$ flow will be strongly stable and so the destabilizing issues associated with elastic effects will not be a concern. Therefore, generally speaking, elastic effects will be detrimental to the manufacturing goals for fiber drawing, but maybe beneficial to the manufacturing goals for extrusion flows with large air holes.

\vspace{0.5cm}
\noindent\textbf{Funding.} J.J.W. was supported by the Research Grants Council of Hong Kong Special Administrative Region, China (CityU 11309422).

\vspace{0.5cm}
\noindent\textbf{Declaration of interests.} The authors report no conflict of interest.

\bibliographystyle{apalike}
\bibliography{jfm}
\end{document}